\documentclass[reprint,amsmath,amssymb,aps,nofootinbib]{revtex4-2}

\usepackage{graphicx}
\usepackage{dcolumn}
\usepackage{bm}

\usepackage{physics}
\usepackage[range-phrase=--]{siunitx}
\usepackage{color}
\usepackage{xcolor}
\definecolor{fak4blue}{RGB}{0, 118, 175/89, 34, 2, 9}
\usepackage[colorlinks=true,citecolor=fak4blue,linkcolor=fak4blue,urlcolor=fak4blue]{hyperref}
\usepackage[noabbrev,nameinlink]{cleveref}
\usepackage{tikz}
\usetikzlibrary{arrows.meta}
\usepackage{svg}
\usepackage{float}
\usepackage{multirow}
\usepackage{booktabs}
\usepackage{csquotes}
\usepackage{mathtools}
\usepackage{orcidlink}

\makeatletter
\newcommand\thefontsize[1]{{#1 The current font size is: \f@size pt\par}}
\makeatother

\DeclareMathOperator{\sinc}{sinc}

\hypersetup{
 pdftitle={Parallel Tempered Metadynamics: Overcoming potential barriers without surfing or tunneling},
 pdfauthor={Eichhorn, Fuwa, Hoelbling, Varnhorst},
 bookmarksnumbered=true,
 bookmarksopen=true,
 bookmarksopenlevel=2
}

\begin{document}

\preprint{WUB/23-00}

\title{Parallel Tempered Metadynamics:\texorpdfstring{\\}{} Overcoming potential barriers without surfing or tunneling}

\author{Timo Eichhorn \orcidlink{0000-0001-9370-36420}}
 \email{timo.eichhorn@protonmail.com}
\author{Gianluca Fuwa \orcidlink{0000-0002-8195-4900}}
 \email{gianluca.fuwa@uni-wuppertal.de}
\author{Christian Hoelbling \orcidlink{0000-0001-5715-1086}}
 \email{hch@uni-wuppertal.de}
\author{Lukas Varnhorst \orcidlink{0000-0002-3718-0143}}
 \email{varnhorst@uni-wuppertal.de}
\affiliation{
 Department of Physics, University of Wuppertal, Gaußstraße 20, 42119 Wuppertal, Germany
}

\date{\today}

\begin{abstract}
At fine lattice spacings, Markov chain Monte Carlo simulations of QCD and other gauge theories with or without fermions are plagued by slow modes that give rise to large autocorrelation times. This can lead to simulation runs that are effectively stuck in one topological sector, a problem known as topological freezing. Here, we demonstrate that for a relevant set of parameters, Metadynamics can be used to unfreeze 4-dimensional SU(3) gauge theory. 
However, compared to local update algorithms and the Hybrid Monte Carlo algorithm, the computational overhead is significant in pure gauge theory, and the required reweighting procedure may considerably reduce the effective sample size. To deal with the latter problem, we propose modifications to the Metadynamics bias potential and the combination of Metadynamics with parallel tempering. We test the new algorithm in 4-dimensional SU(3) gauge theory and find that it can achieve topological unfreezing without compromising the effective sample size, thereby reducing the autocorrelation times of topological observables by at least two orders of magnitude compared to conventional update algorithms. Additionally, we observe significantly improved scaling of autocorrelation times with the lattice spacing in 2-dimensional U(1) gauge theory.
\end{abstract}

\maketitle
\section{Introduction}
\label{sec:1}
In recent years, physical predictions based on lattice simulations have reached sub-percent accuracies \cite{FlavourLatticeAveragingGroupFLAG:2021npn}. With ever-shrinking uncertainties, the demand for precise extrapolations to the continuum grows, which in turn necessitates finer lattice spacings. Current state-of-the-art methods for simulations of lattice gauge theories either rely on a mixture of heat bath \cite{PhysRevLett.43.553, Creutz:1980zw, Fabricius:1984wp, Kennedy:1985nu, Cabibbo:1982zn} and overrelaxation \cite{Adler:1981sn, Creutz:1987xi, Brown:1987rra} algorithms for pure gauge theories or molecular-dynamics-based algorithms like the Hybrid Monte Carlo algorithm (HMC) \cite{Duane:1987de} for simulations including dynamical fermions. For all of these algorithms, the computational effort to carry out simulations dramatically increases at fine lattice spacings due to critical slowing down. While the exact behavior depends on a number of factors, such as the update algorithms, the exact discretization of the action, and the choice of boundary conditions, the scaling of the integrated autocorrelation times with the inverse lattice spacing can usually be described by a power law.

In addition to the general diffusive slowing down, topologically non-trivial gauge theories may exhibit topological freezing \cite{Alles:1996vn, Orginos:2001xa, DelDebbio:2002xa, DeGrand:2002vu, Noaki:2002ai, DelDebbio:2004xh, Aoki:2005ga, Schaefer:2010hu}. This effect appears due to the inability of an algorithm to overcome the action barriers between topological sectors, which can lead to extremely long autocorrelation times of topological observables and thus an effective breakdown of ergodicity.

Over the years, several strategies have been developed to deal with this situation. On the most basic level, it has become customary in large scale simulations to monitor the topological charge of the configurations in each ensemble, thus avoiding regions of parameter space which are affected by topological freezing \cite{BMW:2010skj,Bernard:2017npd,Borsanyi:2020mff}. Another possibility to circumvent the problem consists in treating fixed topology as a finite volume effect and either correcting observables for it \cite{Brower:2003yx,Aoki:2007ka} or increasing the physical volume sufficiently to derive the relevant observables from local fluctuations \cite{Luscher:2017cjh}. It is also possible to use open boundary conditions in one lattice direction \cite{Luscher:2011kk}, which invalidates the concept of an integer topological charge for the price of introducing additional boundary artifacts and a loss of translational symmetry.

Despite the success of these strategies in many relevant situations, the need for a genuine topology changing update algorithm is still great. This is evident from the large number and rather broad spectrum of approaches that are currently being investigated in this direction. Some of these approaches address critical slowing down in general, whereas others focus particularly on topological freezing. These approaches include parallel tempering \cite{Joo:1998ib, Ilgenfritz:2001jp, Borsanyi:2022xml}, modified boundary conditions \cite{Mages:2015scv} and combinations of both \cite{Hasenbusch:2017unr, Bonanno:2020hht}; multiscale thermalization \cite{Endres:2015yca, Detmold:2016rnh, Detmold:2018zgk}, instanton(-like) updates \cite{FUCITO1984230, Smit:1987fq, Dilger:1992yn, Dilger:1994ma, Durr:2012te, Albandea:2021lvl}, Metadynamics \cite{Laio:2015era, Bonati:2017nhe}, multicanonical simulations \cite{PhysRevLett.68.9, Bonati:2018blm}, density of states \cite{Cossu:2021bgn}, Fourier acceleration \cite{DUANE1986143, Duane:1988vr, Davies:1989vh, Cossu:2017eys, Nguyen:2021zgx}, and trivializing maps \cite{Luscher:2009eq, Engel:2011re, Boyle:2022xor}. Additionally, recent years have seen multitudinous efforts to utilize generative models to sample configurations more efficiently \cite{Albergo:2019eim, Kanwar:2020xzo, Nicoli:2020njz, Boyda:2020hsi, DelDebbio:2021qwf, Albergo:2021bna, Hackett:2021idh, Nicoli:2021inv, Foreman:2021ljl, Finkenrath:2022ogg, Albergo:2022qfi, Pawlowski:2022rdn, Gerdes:2022eve, Singha:2022lpi, Abbott:2022zhs, Abbott:2022hkm, Abbott:2022zsh, Bacchio:2022vje, Komijani:2023fzy, Albandea:2023wgd, Nicoli:2023qsl, Abbott:2023thq}. For a recent review of both conventional and machine learning approaches see \cite{Finkenrath:2023sjg}.

In this work we propose a new update algorithm, parallel tempered Metadynamics (PT-MetaD), and demonstrate its efficiency in 4-dimensional SU(3) at parameter values where conventional update algorithms suffer from topological freezing. In its basic variant, which we present here, PT-MetaD consists of two update streams simulating the same physical system. One of the streams uses any suitable, conventional algorithm, while the other one includes a bias potential that facilitates tunneling between topological sectors. At regular intervals, swaps between the two streams are suggested, so that the good topological sampling from the bias potential stream carries over to the other stream. The algorithm thus combines ideas from parallel tempering \cite{PhysRevLett.57.2607}, Metadynamics \cite{Laio_2002}, and multicanonical simulations \cite{PhysRevLett.68.9}, leading to an efficient sampling of topological sectors while avoiding the problem of small effective sample sizes which is usually associated with techniques involving reweighting, such as Metadynamics or multicanonical simulations. Additionally, applying PT-MetaD to theories including fermions is conceptually straightforward.

This paper is organized as follows. We start out by giving a general introduction to Metadynamics in \Cref{sec:2}. Afterwards, \Cref{sec:3} describes our simulation setup, including our choice of actions, observables, and update algorithms. Some details on the application of Metadynamics in the context of SU(3) gauge theory are also given. In \Cref{sec:4}, we present baseline results obtained with conventional update algorithms, including a rough determination of gradient flow scales for the doubly blocked Wilson (DBW2) action. In \Cref{sec:5} we present results obtained with pure Metadynamics for 4-dimensional SU(3) and discuss several possible improvements. In \Cref{sec:6} we introduce parallel tempered Metadynamics and show some scaling tests of the new algorithm in 2-dimensional U(1) gauge theory, as well as exploratory results in 4-dimensional SU(3). Finally, in \Cref{sec:7}, we conclude with a summary and outlook on the application of the new algorithm to full QCD. \Cref{appendix:Conventions,appendix:Clover_charge_derivative,appendix:Cayley-Hamilton_exponential,appendix:Stout_force_recursion} contain mathematical conventions and derivations related to the Metadynamics force calculation, and \Cref{appendix:U1_exact_solution} summarizes exact results for observables in 2-dimensional U(1) gauge theory.
\section{Metadynamics}
\label{sec:2}
Consider a system described by a set of degrees of freedom $\{U\}$, where the states are distributed according to the probability density
\begin{equation}
    p(U) = \frac{1}{Z} e^{-S(U)},
    \label{eq:pU}
\end{equation}
with the partition function $Z$ defined as
\begin{equation}
    Z = \int \mathcal{D}[U] \, e^{-S(U)}.
    \label{eq:Z}
\end{equation}
The expectation value of an observable $O$ is defined as
\begin{equation}
    \langle O \rangle = \int \mathcal{D}[U] \, p(U) O(U).
\end{equation}
In the context of lattice gauge theories, the integration measure $\mathcal{D}[U]$ is usually the product of Haar measures for each link variable, but more generally $\mathcal{D}[U]$ may be understood as a measure on the configuration space of the system.

Metadynamics \cite{Laio_2002} is an enhanced-sampling method, based on the addition of a history-dependent bias potential $V_t(s(U))$ to the action $S(U)$, where $t$ is the current simulation time.
The dynamics of the modified system are governed by $S^M_t(U) = S(U) + V_t(s(U))$, and now explicitly depend on a number of observables $s_i(U)$, with $i \in \{1, \dots, N \}$, that are referred to as collective variables (CVs). These CVs span a low-dimensional projection of the configuration space of the system, and may generally be arbitrary functions of the underlying degrees of freedom $\{U\}$. However, when used in combination with molecular-dynamics-based algorithms, such as the Hybrid Monte Carlo algorithm, the CVs need to be differentiable functions of the underlying degrees of freedom. During the course of a simulation, the bias potential is modified in such a way as to drive the system away from regions of configuration space that have been explored previously, eventually converging towards an estimate of the negative free energy as a function of the CVs, up to a constant offset \cite{PhysRevLett.96.090601, PhysRevE.81.055701}. Usually, this is accomplished by constructing the potential from a sum of Gaussians $g(s)$, so that at simulation time $t$, the potential is given by
\begin{equation}
    V_t(s) = \sum\limits_{t' \leq t} \prod\limits_{i=1}^{N} g\bigl(s_i - s_i(t')\bigr).
\end{equation}
The exact form of the Gaussians is determined by the parameters $w$ and $\delta s_i$:
\begin{equation}
    g(s_i) = w \exp(-\frac{s_i^2}{2 \delta s_i^2}).
\end{equation}
Both parameters affect the convergence behavior of the potential in a similar way: Increasing the height $w$ or the widths $\delta s_i$ may accelerate the convergence of the potential during early stages of the simulation but lead to larger fluctuations around the equilibrium during later stages. Furthermore, the widths $\delta s_i$ effectively introduce a smallest scale that can still be resolved in the space spanned by the CVs, which needs to be sufficiently small to capture the relevant details of the potential.

If the bias potential has reached a stationary state, i.e., its time-dependence in the region of interest is just an overall additive constant, the modified probability density, which we shall also refer to as target density, is given by
\begin{equation}
    p'(U) = \frac{1}{Z'} e^{-S(U) - V(s(U))},
\end{equation}
with the modified partition function
\begin{equation}
    Z' = \int \mathcal{D}[U] \, e^{-S(U) - V(s(U))}.
\end{equation}
Expectation values with respect to the modified distribution can then be defined in the usual way, i.e., via
\begin{equation}
    \langle O \rangle' = \int \mathcal{D}[U] \, p'(U) O(U).
\end{equation}
On the other hand, expectation values with respect to the original, unmodified probability density can be written in terms of the modified probability distribution with an additional weighting factor. For a dynamic potential, different reweighting schemes have been put forward to achieve this goal \cite{doi:10.1021/acs.jctc.9b00867}, but if the potential is static, the weighting factors are directly proportional to the exponential of the bias potential:
\begin{equation}
    \langle O \rangle = \frac{\displaystyle\int \mathcal{D}[U] \, p'(U) O(U) \, e^{V(s(U))}}{\displaystyle\int \mathcal{D}[U] \, e^{V(s(U))}}.
\end{equation}
The case of a static potential is thus essentially the same as a multicanonical simulation \cite{PhysRevLett.68.9}.

In situations where the evolution of the system is hindered by high action barriers separating relevant regions of configuration space, Metadynamics can be helpful in overcoming those barriers, since the introduction of a bias potential modifies the marginal distribution over the set of CVs. For conventional Metadynamics, the bias potential is constructed in such a way that the marginal modified distribution is constant:
\begin{equation}
    p'(s_i) = \int \mathcal{D}[U] \, p'(U) \delta\bigl(s_i - s_i(U)\bigr) = \mathrm{const.}
\end{equation}
Conversely, for a given original distribution $p(s)$ and a desired target distribution $p'(s)$, the required potential is given by:
\begin{equation}
    V(s) = \ln(\frac{p'(s)}{p(s)}).
\end{equation}
Nevertheless, it is important to recognize that even if the bias potential completely flattens out the marginal distribution over the CVs, the simulation is still expected to suffer from other (diffusive) sources of critical slowing down as is common for Markov chain Monte Carlo simulations.
\section{Simulation setup and observables}
\label{sec:3}
\subsection{Choice of gauge actions}
\label{subsec:3.1}
For our simulations of SU(3) gauge theory, we work on a 4-dimensional lattice $\Lambda$ with periodic boundary conditions. Configurations are generated using the Wilson \cite{Wilson:1974sk} and the DBW2 \cite{Takaishi:1996xj} gauge actions, both of which belong to a one-parameter family of gauge actions involving standard $1 \times 1$ plaquettes as well as $1 \times 2$ planar loops, which may be expressed as
\begin{equation}
    \begin{aligned}
        S_g = \frac{\beta}{3} \sum\limits_{n \in \Lambda} \Bigl( \sum\limits_{\mu < \nu} &c_0 \bigl( 3 - \Re\tr[\mathcal{W}_{\mu, \nu}(n)] \bigr) \\
            + \sum\limits_{\mu \neq \nu} &c_1 \bigl( 3 - \Re\tr[\mathcal{W}_{\mu, 2\nu}(n)] \bigr) \Bigr).
    \end{aligned}
\end{equation}
Here, $\mathcal{W}_{k \mu, l \nu}(n)$ refers to a Wilson loop of shape $k \times l$ in the $\mu$-$\nu$ plane originating at the site $n$. The coefficients $c_0$ and $c_1$ are constrained by the normalization condition $c_0 + 8 c_1 = 1$ and the positivity condition $c_0 > 0$, where the latter condition is sufficient to guarantee that the set of configurations with minimal action consists of locally pure gauge configurations \cite{Luscher:1984xn}. For the Wilson gauge action ($c_1 = 0$), only plaquette terms contribute, whereas the DBW2 action ($c_1 = -1.4088$) also involves rectangular loops.

It is well known that the critical slowing down of topological modes is more pronounced for improved gauge actions than for the Wilson gauge action \cite{Orginos:2001xa, DeGrand:2002vu, Noaki:2002ai, Aoki:2005ga, Schaefer:2010hu}: A larger negative coefficient $c_1$ suppresses small dislocations, which are expected to be the usual mechanism mediating transitions between topological sectors on the lattice. Among the most commonly used gauge actions, this effect is most severely felt by the DBW2 action. In previous works \cite{Noaki:2002ai, Aoki:2005ga}, local update algorithms were found to be inadequate for exploring different topological sectors in a reasonable time frame at lattice spacings around \SI{0.06}{\femto\meter}. Instead, the authors had to generate thermalized configurations in different topological sectors using the Wilson gauge action, before using these configurations as starting points for simulations with the DBW2 action. Thus, this action allows us to explore parameters where severe critical slowing down is visible, while avoiding very fine lattice spacings and thereby limiting the required computational resources.

\subsection{Observables}
\label{subsec:3.2}
The observables we consider here are based on various definitions of the topological charge, and Wilson loops of different sizes at different smearing levels. The unrenormalized topological charge is defined using the clover-based definition of the field strength tensor:
\begin{equation}
    Q_c = \frac{1}{32 \pi^2} \sum_{n \in \Lambda} \epsilon_{\mu \nu \rho \sigma} \tr[\Hat{F}^{\mathrm{clov}}_{\mu \nu}(n) \Hat{F}^{\mathrm{clov}}_{\rho \sigma}(n)].
    \label{eq:top_charge_clov_def}
\end{equation}
This field strength tensor is given by
\begin{equation}
    \Hat{F}^{\mathrm{clov}}_{\mu \nu}(n) = -\frac{i}{8} \bigl(C_{\mu \nu}(n) - C_{\nu \mu}(n) \bigr),
\end{equation}
where the clover term $C_{\mu \nu}(n)$ is defined as
\begin{equation}
    C_{\mu \nu}(n) = P_{\mu, \nu}(n) + P_{\nu, -\mu}(n) + P_{-\mu, -\nu}(n) + P_{-\nu, \mu}(n),
\end{equation}
and $P_{\mu, \nu}(n)$ denotes the plaquette:
\begin{equation}
    P_{\mu, \nu}(n) = U_{\mu}(n) U_{\nu}(n + \hat{\mu}) U_{\mu}^{\dagger}(n + \hat{\nu}) U_{\nu}^{\dagger}(n).
\end{equation}
Alternatively, the topological charge may also be defined via the plaquette-based definition, here denoted by $Q_p$:
\begin{equation}
    Q_p = \frac{1}{32 \pi^2} \sum_{n \in \Lambda} \epsilon_{\mu \nu \rho \sigma} \tr[\Hat{F}^{\mathrm{plaq}}_{\mu \nu}(n) \Hat{F}^{\mathrm{plaq}}_{\rho \sigma}(n)].
    \label{eq:top_charge_plaq_def}
\end{equation}
Similar to the clover-based field strength tensor, the plaquette-based field strength tensor is defined as:
\begin{equation}
    \Hat{F}^{\mathrm{plaq}}_{\mu \nu}(n) = -\frac{i}{2} \bigl(P_{\mu, \nu}(n) - P_{\nu, \mu}(n) \bigr).
\end{equation}
Both $Q_c$ and $Q_p$ formally suffer from $\mathcal{O}(a^2)$ artifacts, although the coefficient is typically smaller for the clover-based definition $Q_c$. The topological charge is always measured after $\mathcal{O}(30)$ steps of stout smearing \cite{Morningstar:2003gk} with a smearing parameter $\rho = 0.12$. To estimate the autocorrelation times of the system, we consider the squared topological charge \cite{Schaefer:2010hu}, the Wilson gauge action, and $n \times n$ Wilson loops for $n \in \{2, 4, 8\}$ at different smearing levels. We denote the latter two by $S_{w}$ and $\mathcal{W}_{n}$, respectively.

\subsection{Update algorithms}
\label{subsec:3.3}
Throughout this work, we employ a number of different update schemes: To illustrate critical slowing down of conventional update algorithms and to set a baseline for comparison with Metadynamics-based algorithms, we use standard Hybrid Monte Carlo updates with unit length trajectories (1HMC), a single heat bath sweep (1HB), five heat bath sweeps (5HB), and a single heat bath sweep followed by four overrelaxation sweeps (1HB+4OR). The local update algorithms are applied to three distinct SU(2) subgroups during each sweep \cite{Cabibbo:1982zn}, and the HMC updates use an Omelyan-Mryglod-Folk fourth-order minimum norm integrator \cite{OMELYAN2003272} with a step size of $\epsilon = 0.2$, which leads to acceptance rates above \SI{99}{\percent} for the parameters used here.

We compare these update schemes to Metadynamics HMC updates with unit length trajectories (MetaD-HMC), and a combination of parallel tempering with Metadynamics (PT-MetaD) which is discussed in more detail in \Cref{sec:6}.

An important requirement for the successful application of Metadynamics is the identification of appropriate CVs. In our case, the CV should obviously be related to the topological charge. However, it should not always be (close to) integer-valued but rather reflect the geometry of configuration space with respect to the boundaries between topological sectors. On the other hand, the CV needs to track the topological charge closely enough for the algorithm to be able to resolve and overcome the action barriers between topological sectors. A straightforward approach is to apply only a moderate amount of some kind of smoothing procedure, such as cooling or smearing, to a gluonic definition of the topological charge, for which we choose $Q_c$. Since these smoothing procedures involve spatial averaging, the action will become less local, which complicates the use of local update algorithms. Therefore, we use the HMC algorithm to efficiently update the entire gauge field at the same time, which requires a differentiable smoothing procedure, such as stout \cite{Morningstar:2003gk} or HEX smearing \cite{Capitani:2006ni}. Due to its simpler implementation compared to HEX smearing, we choose stout smearing here. Previous experience \cite{Eichhorn:2022wxn} seems to indicate that for $Q_c$, four to five stout smearing steps with a smearing parameter $\rho = 0.12$ strike a reasonable balance between having a smooth CV and still representing the topological charge accurately. We found that using $Q_p$ would require significantly more smearing steps, whereas some improved definitions involving more general rectangular loops would not reduce the necessary amount of smearing.

The force contributed by the bias potential may be written in terms of the chain rule:
\begin{equation}
    \begin{aligned}
        F_{\mu, \mathrm{meta}}(n) = &-\frac{\partial V_{\mathrm{meta}}}{\partial Q_{\mathrm{meta}}} \frac{\partial Q_{\mathrm{meta}}}{\partial U^{(s)}_{\mu_s}(n_s)} \\
                                    &\times\frac{\partial U^{(s)}_{\mu_s}(n_s)}{\partial U^{(s-1)}_{\mu_{s-1}}(n_{s-1})} \dots \frac{\partial U^{(1)}_{\mu_1}(n_1)}{\partial U_{\mu}(n)}.
        \label{eq:metadynamics_force}
    \end{aligned}
\end{equation}
Here we have introduced the notation $V_{\mathrm{meta}}$ for the bias potential and $Q_{\mathrm{meta}}$ for the CV to clearly distinguish it from other definitions of the topological charge. Note that there is an implicit summation over the lattice sites $n_i$ and the Lorentz indices $\mu_i$. The first term in the equation, corresponding to the derivative of the bias potential with respect to $Q_{\mathrm{meta}}$, is trivial, but the latter two terms are more complicated: The derivative of $Q_{\mathrm{meta}}$ with respect to the maximally smeared field $U^{(s)}$ is given by a sum of staples with clover term insertions, and the final terms correspond to the stout force recursion \cite{Morningstar:2003gk} that also appears during the force calculation when using smeared fermions. Note that in machine learning terminology, this operation is essentially a backpropagation \cite{Nagai:2021bhh} and may be computed efficiently using reverse mode automatic differentiation. More details on the calculation of the force can be found in \Cref{appendix:Clover_charge_derivative,appendix:Stout_force_recursion,appendix:Cayley-Hamilton_exponential}.

The bias potential is constructed from a sum of one-dimensional Gaussians, as described in \Cref{sec:2}, and stored as a histogram. Due to charge conjugation symmetry, we can update the potential symmetrically. Values at each point are reconstructed by linearly interpolating between the two nearest bins, and the derivative is approximated by their finite difference. To limit the evolution to relevant regions of the phase space, we introduce an additional penalty term to the potential once the value of $Q_{\mathrm{meta}}$ has crossed certain thresholds $Q_{\mathrm{min}}$ and $Q_{\mathrm{max}}$. If the system has exceeded the threshold, the potential is given by the outermost value of the histogram, plus an additional term that scales quadratically with the distance to the outer limit of the histogram.

Unless mentioned otherwise, we have used the following values as default parameters for the potential: $Q_{\mathrm{max/min}} = \pm8$, $n_{\mathrm{bins}} = 800$, $w = 0.05$, while $\delta Q^2$ has always been set equal to the bin width, i.e., $(Q_{\mathrm{max}} - Q_{\mathrm{min}}) / n_{\mathrm{bins}}$.

It is often convenient to build up a bias potential in one or several runs, and then simulate and measure with a static potential generated from the previous runs. In some sense, this can be thought of as a combination of Metadynamics and multicanonical simulations.
\section{Results with conventional update algorithms}
\label{sec:4}
To establish a baseline to compare our results to, we have investigated the performance of some conventional update algorithms using the Wilson and DBW2 gauge actions. Furthermore, we have made a rough determination of the gradient flow scales $t_0$ and $w_0$ for the DBW2 action. Some preliminary results for the Wilson action were already presented in \cite{Eichhorn:2022wxn}.
\subsection{Critical slowing down with Wilson and DBW2 gauge actions}
\label{subsec:4.1}
In order to study the scaling of autocorrelations for different update schemes, we have performed a series of simulations with the Wilson gauge action on a range of lattice spacings. The parameters were chosen to keep the physical volume approximately constant at around $(\SI{1.1}{\femto\meter})^4$, using the scale given by the rational fit function in \cite{Durr:2006ky}, which was based on data from \cite{Necco:2001xg}. A summary of the simulation parameters can be found in \Cref{tab:wilson_conventional_parameters}.
\begin{table}[h]
\caption{A summary of the simulation parameters for the Wilson gauge action runs using conventional update algorithms. The scale was set via the rational fit from \cite{Durr:2006ky} (where $r_0 = \SI{0.49}{\femto\meter}$), which in turn used data from \cite{Necco:2001xg}.}
\label{tab:wilson_conventional_parameters}
\begin{ruledtabular}
    \begin{tabular}{cccc}
         $\beta$ & $L/a$ & $a$ [fm] & $N_{\mathrm{conf}}$ \\ \toprule
         5.8980  & $10$  & $0.1097$ & $100000$            \\
         6.0000  & $12$  & $0.0914$ & $100000$            \\
         6.0938  & $14$  & $0.0783$ & $100000$            \\
         6.1802  & $16$  & $0.0686$ & $100000$            \\
         6.2602  & $18$  & $0.0610$ & $100000$            \\
         6.3344  & $20$  & $0.0549$ & $100000$            \\
         6.4035  & $22$  & $0.0499$ & $100000$
    \end{tabular}
\end{ruledtabular}
\end{table}

\begin{figure*}
    \centering
    \includegraphics{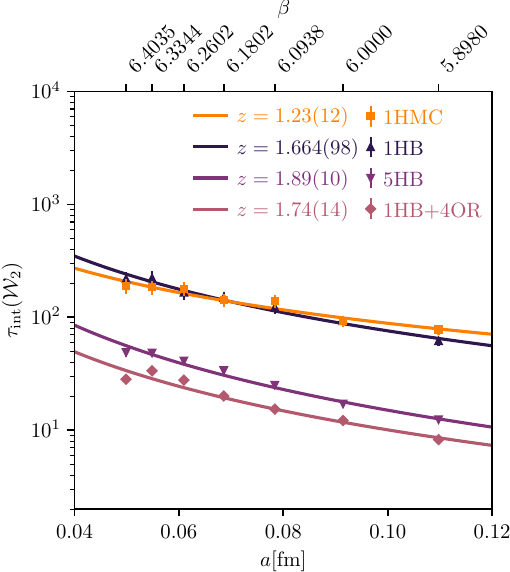}
    \includegraphics{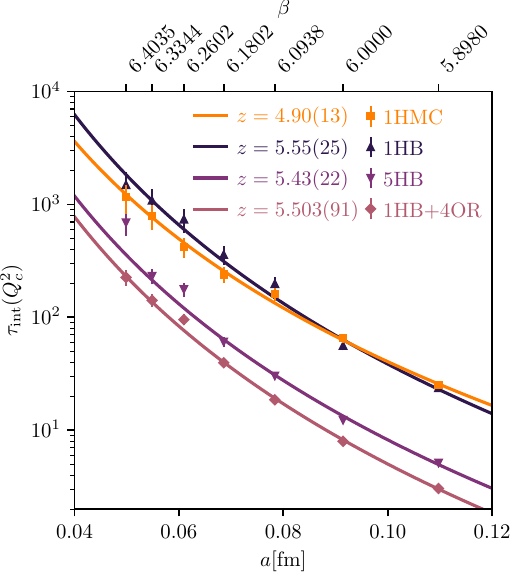}
    \caption{Scaling of the integrated autocorrelation times of square $2 \times 2$ Wilson loops $\mathcal{W}_{2}$ (left) and the squared topological charge $Q_c^2$ (right) for different update schemes using the Wilson gauge action. The scaling of both observables can be described using a power-law fit (\Cref{eq:autocorrelation_power_law_fit}) and is compatible with a dynamical critical exponent $z \approx 2$ for the Wilson loops and $z \approx 5$ for the squared topological charge. Details on the simulation parameters are listed in \Cref{tab:wilson_conventional_parameters}.}
    \label{fig:tau_int_scaling_wilson}
\end{figure*}

Since autocorrelation times near second-order phase transitions are expected to be described by a power law, we use the following fit ansatz in an attempt to parametrize the scaling:
\begin{equation}
     \tau_{\mathrm{int}} = c \biggl(\frac{r_0}{a}\biggr)^z
     \label{eq:autocorrelation_power_law_fit}
\end{equation}
All autocorrelation times and their uncertainties are estimated following the procedure described in \cite{Wolff:2003sm}.
\Cref{fig:tau_int_scaling_wilson} shows the scaling of the integrated autocorrelation times of $2 \times 2$ Wilson loops $\mathcal{W}_2$ and the square $Q_c^2$ of the clover-based topological charge with the lattice spacing. Additionally, the figure also includes power law fits to the data and the resulting values for the dynamical critical exponents $z(\mathcal{W}_2)$ and $z(Q_c^2)$. Both observables were measured after 31 stout smearing steps with a smearing parameter $\rho = 0.12$.
While the integrated autocorrelation times of both observables increase towards finer lattice spacings and are adequately described by a power law behavior, the increase is much steeper for the squared topological charge than for the smeared $2 \times 2$ Wilson loops. Below a crossover point at $a \approx \SI{0.08}{\femto\meter}$, the autocorrelation times of the squared topological charge start to dominate. They can be described by both a dynamical critical exponent $z \approx 5$ or, alternatively, by an exponential increase that was first suggested in \cite{DelDebbio:2004xh}. This behavior is compatible with the observations in \cite{Schaefer:2010hu}.

In contrast, the autocorrelation time of Wilson loops is compatible with a much smaller exponent $z \approx$ \numrange{1}{2}. As can be seen in \Cref{tab:dynamical_critical_exponents_wilson}, the critical exponent does not change significantly with the size of the Wilson loop after 31 stout smearing steps. Generally, the integrated autocorrelation times of smeared Wilson loops slightly increase both with the size of the loops and the number of smearing levels. The only exception to this behavior occurs for larger loops, where a few steps of smearing are required to obtain a clean signal and not measure the autocorrelation of the noise instead.

Regarding the different update algorithms, the unit length HMC does show a somewhat better scaling behavior for all observables than the local update algorithms, but is also the most computationally expensive update scheme considered here (see \Cref{tab:update_times}).\footnote{Since we are ultimately interested in dynamical fermion simulations, we do not consider the more efficient local HMC variant presented in \cite{Marenzoni:1993im}, as it is applicable to local bosonic actions only.} For all the local update algorithms, the critical exponents are very similar, but the combination of one heat bath and four overrelaxation steps has the smallest prefactor. It is interesting to note that this algorithm is also approximately twice as fast as the five-step heat bath update scheme, while still providing smaller autocorrelation times. The single step heat bath without overrelaxation, although numerically cheapest, exhibits the worst prefactor of the local update algorithms. Note that the reported numbers for the critical exponents differ from those in \cite{Eichhorn:2022wxn} due to a different fit ansatz (in the proceedings, the ansatz included an additional constant term).
\begin{table}[h]
\caption{Relative performance of the update algorithms used in our scaling runs. The results cited here are taken from simulations on $22^4$ lattices. Note that the performance of the heat bath algorithm is slightly better for larger $\beta$ \cite{Fabricius:1984wp, Kennedy:1985nu}.}
\label{tab:update_times}
\begin{ruledtabular}
    \begin{tabular}{cc}
         Update scheme & Relative time \\ \toprule
         1HMC          & 6.98          \\
         1HB           & 1.00          \\
         5HB           & 4.99          \\
         1HB+4OR       & 2.02          \\
    \end{tabular}
\end{ruledtabular}
\end{table}
\begin{table}[h]
\caption{The dynamical critical exponents obtained from power law fits to the integrated autocorrelation times of $Q_{c}^{2}$, $S_w$, and Wilson loops of different sizes after 31 stout smearing steps for different update schemes and the Wilson gauge action. Notably, the dynamical critical exponents associated with $Q_c^2$ are much larger than those associated with the smeared action or smeared Wilson loops of different sizes.}
\label{tab:dynamical_critical_exponents_wilson}
\begin{ruledtabular}
    \begin{tabular}{cccccc}
        Update  & \multirow{2}{*}{$z(Q_{c}^{2})$} & \multirow{2}{*}{$z(S_w)$} & \multirow{2}{*}{$z(\mathcal{W}_{2})$} & \multirow{2}{*}{$z(\mathcal{W}_{4})$} & \multirow{2}{*}{$z(\mathcal{W}_{8})$} \\
        scheme  &                                 &                           &                                       &                                       &                                       \\ \toprule
        1HMC    & \num{4.90 \pm 0.13}             & \num{1.27 \pm 0.12}       & \num{1.23 \pm 0.12}                   & \num{1.16 \pm 0.12}                   & \num{1.29 \pm 0.16}                   \\
        1HB     & \num{5.55 \pm 0.25}             & \num{1.69 \pm 0.10}       & \num{1.66 \pm 0.10}                   & \num{1.64 \pm 0.09}                   & \num{1.82 \pm 0.12}                   \\
        5HB     & \num{5.43 \pm 0.22}             & \num{1.92 \pm 0.11}       & \num{1.89 \pm 0.10}                   & \num{1.85 \pm 0.10}                   & \num{1.95 \pm 0.10}                   \\
        1HB+4OR & \num{5.50 \pm 0.09}             & \num{1.77 \pm 0.15}       & \num{1.74 \pm 0.14}                   & \num{1.71 \pm 0.14}                   & \num{1.85 \pm 0.13}                   \\
    \end{tabular}
\end{ruledtabular}
\end{table}

\begin{figure}[h]
    \centering
    \includegraphics{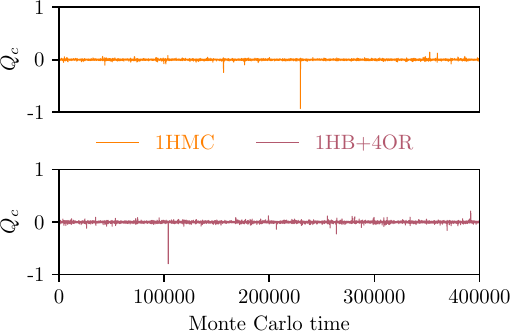}
    \caption{Time series of the topological charge for $V = 16^4$, $\beta = 1.25$ using the DBW2 action. The configurations were generated with the 1HMC (top) and the 1HB+4OR (bottom) update schemes. Out of a total of 400000 configurations each, only a single configuration during the 1HB+4OR run and two (successive) configurations during the 1HMC run fulfill the condition $\abs{Q_c} > 0.5$.}
    \label{fig:timeseries_conventional}
\end{figure}

For the DBW2 action, the problem is more severe. \Cref{fig:timeseries_conventional} shows the time series of the topological charge for two runs using the 1HMC and the 1HB+4OR update scheme. Both simulations were carried out on $16^4$ lattices at $\beta = 1.25$ using the DBW2 action.

Evidently, both update schemes are unable to tunnel between different topological sectors in a reasonable time. Only a single configuration during the 1HB+4OR run and two (successive) configurations during the 1HMC run fulfill the condition $\abs{Q_c} > 0.5$.

\subsection{Scale setting for the DBW2 action}
\label{subsec:4.2}
To the best of our knowledge, scales for the DBW2 action in pure gauge theory have only been computed based on simulations with $\beta \leq 1.22$ \cite{Hashimoto:2004rs, Aoki:2005ga}, and interpolation formulas are only available based on data with $\beta \leq 1.04$ \cite{Necco:2003vh}. Since here we perform simulations at $\beta = 1.25$, we compute approximate values for $t_0$ \cite{Luscher:2010iy} and $w_0$ \cite{BMW:2012hcm}, which allows us to estimate our lattice spacings for comparison to the Wilson results. Both scales are based on the energy density $E$, which is defined as:
\begin{equation}
    \begin{aligned}
        E &= \frac{1}{4a^2V} \sum\limits_{n \in \Lambda} F_{\mu \nu}^{a}(n) F_{\mu \nu}^{a}(n)
        \\&= -\frac{1}{2a^2V} \sum\limits_{n \in \Lambda} \tr[F_{\mu \nu}(n) F_{\mu \nu}(n)].
    \end{aligned}
\end{equation}
Similar to the topological charge definitions, we adopt a plaquette- and clover-based definition of the field strength tensor, with the only difference being that the components are also made traceless, and not just anti-Hermitian. The gradient flow scales $t_0$ and $w_0$ are both defined implicitly:
\begin{alignat}{2}
    \mathcal{E}(t_0) = t^2 \langle E \rangle &\Big\vert_{t = t_0} &= 0.3,
    \\ W(w^2_0) = t \frac{\mathrm{d}}{\mathrm{d}t} \mathcal{E}(t) &\Big\vert_{t = w_0^2} &= 0.3.
\end{alignat}
The flow equation was integrated using the third-order commutator free Runge-Kutta scheme from \cite{Luscher:2010iy} with a step size of $\epsilon = 0.025$. Measurements of the clover-based energy density were performed every 10 integration steps, and $t^2 \langle E(t) \rangle$ was fitted with a cubic spline, which was evaluated with a step size of 0.001. For every value of $\beta$, two independent simulations with 100 measurements each were performed on $48 \times 32^3$ lattices. Every measurement was separated by 200 update sweeps with the previously described 1HB+4OR update scheme, and the initial 2000 updates were discarded as thermalization phase. Our results are displayed in \Cref{tab:flow_scales_dbw2}.
\begin{table}[h]
\caption{Results for different gradient flow scales for the DBW2 gauge action. These results should not be interpreted as an attempt at a precise scale determination but rather as an approximate estimate.}
\label{tab:flow_scales_dbw2}
\begin{ruledtabular}
    \begin{tabular}{cccccc}
        $\beta$ & $N_t \times N_s^3$   & $t_{0, \mathrm{plaq}}/a^2$ & $t_{0, \mathrm{clov}}/a^2$ & $w^2_{0, \mathrm{plaq}}/a^2$ & $w^2_{0, \mathrm{clov}}/a^2$ \\ \toprule
        1.04    & $48\times 32^3$      & \num{3.445 \pm 0.003}      & \num{3.647 \pm 0.003}      & \num{3.601 \pm 0.004}        & \num{3.641 \pm 0.004}      \\
        1.10    & $48\times 32^3$      & \num{4.483 \pm 0.006}      & \num{4.684 \pm 0.006}      & \num{4.675 \pm 0.009}        & \num{4.716 \pm 0.009}      \\
        1.15    & $48\times 32^3$      & \num{5.549 \pm 0.009}      & \num{5.751 \pm 0.010}      & \num{5.787 \pm 0.014}        & \num{5.827 \pm 0.014}      \\
        1.16    & $48\times 32^3$      & \num{5.761 \pm 0.009}      & \num{5.962 \pm 0.009}      & \num{5.992 \pm 0.015}        & \num{6.032 \pm 0.015}      \\
        1.17    & $48\times 32^3$      & \num{6.032 \pm 0.008}      & \num{6.234 \pm 0.008}      & \num{6.291 \pm 0.014}        & \num{6.332 \pm 0.013}      \\
        1.18    & $48\times 32^3$      & \num{6.269 \pm 0.013}      & \num{6.470 \pm 0.014}      & \num{6.525 \pm 0.024}        & \num{6.566 \pm 0.024}      \\
        1.19    & $48\times 32^3$      & \num{6.524 \pm 0.010}      & \num{6.726 \pm 0.010}      & \num{6.803 \pm 0.012}        & \num{6.844 \pm 0.012}      \\
        1.20    & $48\times 32^3$      & \num{6.798 \pm 0.015}      & \num{7.000 \pm 0.015}      & \num{7.082 \pm 0.020}        & \num{7.123 \pm 0.020}      \\
        1.21    & $48\times 32^3$      & \num{7.047 \pm 0.016}      & \num{7.248 \pm 0.016}      & \num{7.331 \pm 0.025}        & \num{7.372 \pm 0.025}      \\
        1.22    & $48\times 32^3$      & \num{7.386 \pm 0.023}      & \num{7.588 \pm 0.024}      & \num{7.710 \pm 0.035}        & \num{7.751 \pm 0.035}      \\
        1.23    & $48\times 32^3$      & \num{7.642 \pm 0.023}      & \num{7.844 \pm 0.024}      & \num{7.954 \pm 0.035}        & \num{7.995 \pm 0.035}      \\
        1.24    & $48\times 32^3$      & \num{7.963 \pm 0.023}      & \num{8.165 \pm 0.023}      & \num{8.293 \pm 0.035}        & \num{8.334 \pm 0.035}      \\
        1.25    & $48\times 32^3$      & \num{8.312 \pm 0.027}      & \num{8.515 \pm 0.028}      & \num{8.681 \pm 0.037}        & \num{8.721 \pm 0.037}
    \end{tabular}
\end{ruledtabular}
\end{table}

Using the physical value of $\sqrt{t_0} = \SI{0.1638 \pm 0.0010}{\femto\meter}$ from \cite{Sommer:2014mea}, these results imply a physical volume of approximately $(\SI{0.9}{\femto\meter})^4$ and a temperature of around \SI{219}{\MeV} for the $16^4$ lattice from the previous section.

\begin{figure}[h]
    \centering
    \includegraphics{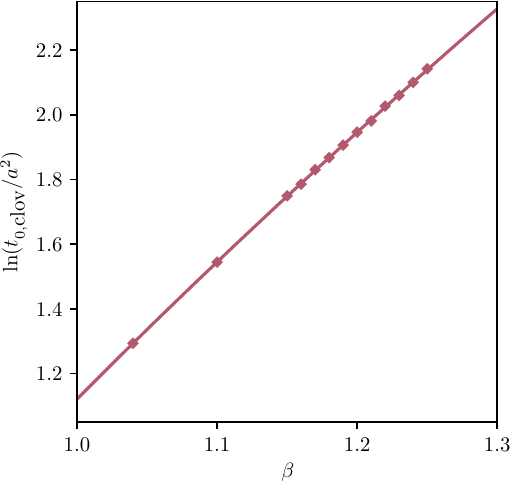}
    \caption{Rational fit of the form \Cref{eq:ln_t0_vs_beta_fit} to the $t_{0, \mathrm{clov}}/a^2$ values presented in \Cref{tab:flow_scales_dbw2}. The fit has $\chi^2 / \mathrm{d.o.f.} \approx 1.31$ and parameters $d_1 \approx 1.0351$, $d_2 \approx -1.3763$, and $d_3 \approx 0.4058$. Error bars are substantially smaller than the symbols.}
    \label{fig:t0_clov_fit_dbw2}
\end{figure}

In order to facilitate comparison with other results, we also provide an interpolation of our lattice spacing results. For this purpose, we use a rational fit ansatz with three fit parameters that is asymptotically consistent with perturbation theory \cite{Durr:2006ky} and has a sufficient number of degrees of freedom to describe our data well:
\begin{equation}
    \ln(t_0 / a^2) = \frac{8 \pi^2}{33} \beta \frac{1 + d_1 / \beta + d_2 / \beta^2}{1 + d_3 / \beta}.
    \label{eq:ln_t0_vs_beta_fit}
\end{equation}
For our reference, clover-based $t_0$ scale setting, this results in a fit with $\chi^2 / \mathrm{d.o.f.} \approx 1.31$ and parameters $d_1 \approx 1.0351$, $d_2 \approx -1.3763$, $d_3 \approx 0.4058$, which is displayed in \Cref{fig:t0_clov_fit_dbw2}.
We want to emphasize that these results are not meant to be an attempt at a precise scale determination but rather only serve as an approximate estimate. Especially for the finer lattices, the proper sampling of the topological sectors cannot be guaranteed, and the comparatively small volumes may introduce non-negligible finite volume effects.
\section{Results with Metadynamics}
\label{sec:5}
\begin{figure*}
    \centering
    \includegraphics{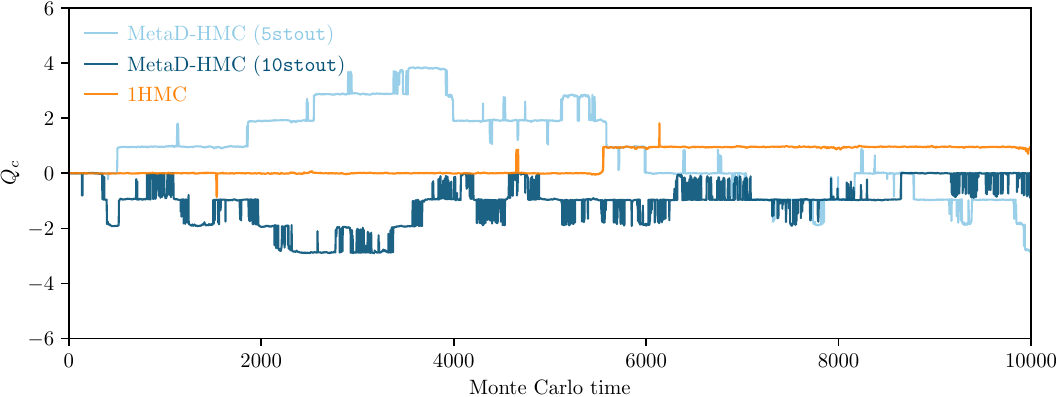}
    \caption{Comparison of the time series of the topological charge between runs using the HMC algorithm and MetaD-HMC runs for $V = 22^4$ and $\beta = 6.4035$ using the Wilson gauge action. The bias potentials in the Metadynamics runs were built up dynamically/from scratch during each run. Both Metadynamics runs are able to transition between topological sectors numerous times, whereas the run using the conventional HMC is essentially stuck in two sectors.}
    \label{fig:timeseries_metad_wilson}
\end{figure*}

\Cref{fig:timeseries_metad_wilson} shows the time series of the topological charge obtained from simulations with the HMC and the MetaD-HMC with five and ten stout smearing steps on a $22^4$ lattice at $\beta = 6.4035$ using the Wilson gauge action.
\begin{figure*}
    \centering
    \includegraphics{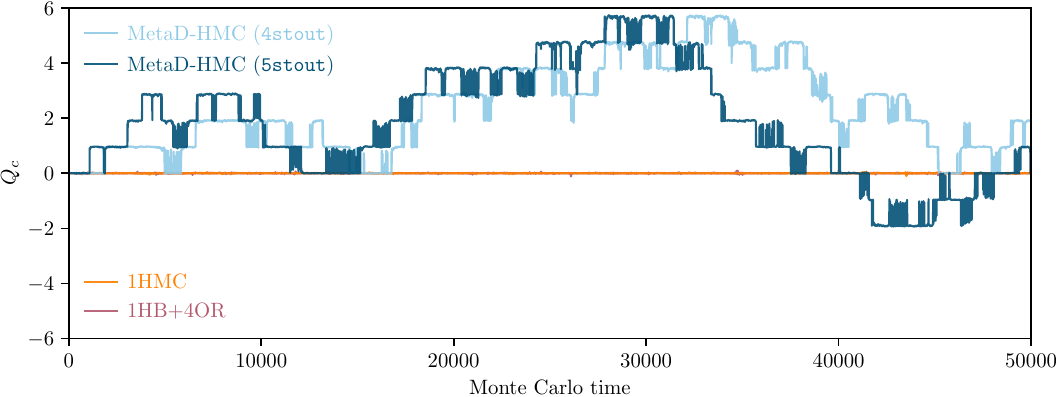}
    \caption{Comparison of the time series of the topological charge between runs using conventional update algorithms (HMC and a combination of heat bath and overrelaxation updates) and MetaD-HMC runs for $V = 16^4$ and $\beta = 1.25$ using the DBW2 action. The bias potentials in the Metadynamics runs were built up dynamically during each run. The results shown for the 1HMC and 1HB+4OR update schemes are from the same runs as the time series shown in \Cref{fig:timeseries_conventional}. While the conventional update algorithms are unable to escape the $Q = 0$ sector, both Metadynamics runs frequently transition between different topological sectors.}
    \label{fig:timeseries_metad_dbw2}
\end{figure*}
Both MetaD-HMC runs tunnel multiple times between different topological sectors, whereas the conventional HMC run essentially displays only a single tunneling event between sectors $Q = 0$ and $Q = 1$. The autocorrelation times of $Q_c^2$ for the MetaD-HMC runs are comparable despite the different amounts of smearing used to define the CV: For the run with five smearing steps $\tau_{\mathrm{int}}(Q_c^2) = \num{586 \pm 142}$, while $\tau_{\mathrm{int}}(Q_c^2) = \num{342 \pm 68}$ for the run with ten smearing steps. A noteworthy difference between the two MetaD-HMC runs is the increase of fluctuations with higher amounts of smearing. If too many smearing steps are used to define the CV, the resulting $Q$ values will generally be closer to integers, which will eventually drive the system to coarser regions of configuration space. Since these regions do not contribute significantly to expectation values in the path integral, it is desirable to minimize the time that the algorithm spends there. This is directly related to the issue of small effective sample sizes after reweighting, which we will discuss in more detail in \Cref{subsec:5.2}.

A similar comparison of topological charge time series for the DBW2 action can be seen in \Cref{fig:timeseries_metad_dbw2}. Here, two MetaD-HMC runs with four and five stout smearing steps on a $16^4$ lattice at $\beta = 1.25$ are compared to the 1HMC and 1HB+4OR runs, which were already shown in \Cref{fig:timeseries_conventional}. Both conventional update schemes are confined to the zero sector, whereas the two MetaD-HMC runs explore topological sectors up to $\abs{Q} = 6$. More quantitatively, the integrated autocorrelation time of $Q_c^2$ is estimated to be $\tau_{\mathrm{int}}(Q_c^2) = \num{5126 \pm 1500}$ for the run with 4 smearing steps and $\tau_{\mathrm{int}}(Q_c^2) = \num{4159 \pm 1137}$ for the run with 5 smearing steps. On the other hand, lower bounds for the autocorrelation times of the 1HMC and 1HB+4OR update schemes are $\num{4e5}$, which is larger by more than a factor of 70.

\begin{figure*}
    \centering
    \includegraphics{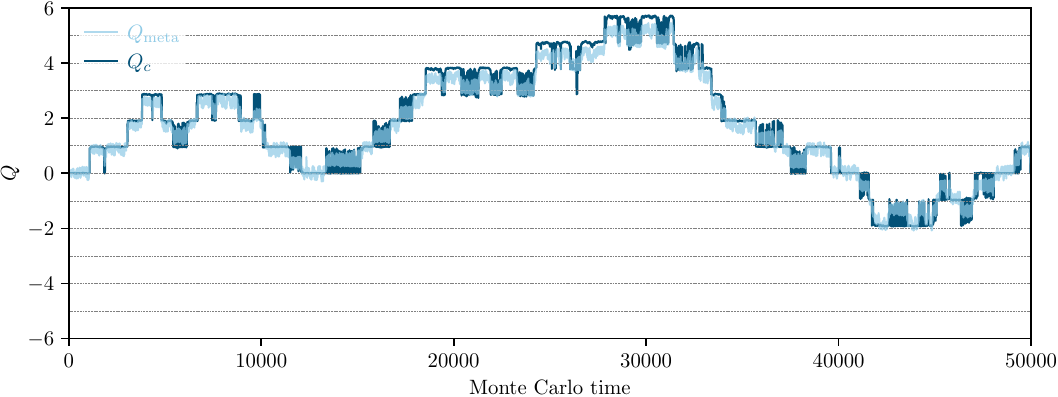}
    \caption{Time series of the CV $Q_{\mathrm{meta}}$ and the topological charge $Q_c$, measured after 5 and 30 stout smearing steps with a smearing parameter of $\rho = 0.12$, respectively. The data are from the \texttt{5stout} MetaD-HMC run shown in \Cref{fig:timeseries_metad_dbw2}.}
    \label{fig:timeseries_metacharge_topcharge_dbw2}
\end{figure*}

To illustrate the role of the CV $Q_{\mathrm{meta}}$, it may be helpful to compare the time series of $Q_{\mathrm{meta}}$ and $Q_{c}$, as shown in \Cref{fig:timeseries_metacharge_topcharge_dbw2}.
The two observables are clearly correlated, but $Q_\mathrm{meta}$ is distributed more evenly between integers. 

\subsection{Computational overhead and multiple timescale integration}
\label{subsec:5.1}
A fair comparison of the different update schemes also needs to take the computational cost of the algorithms into account. \Cref{tab:algorithm_benchmarks} shows the relative timings for the different update schemes used in this section, measured for simulations carried out on $16^4$ lattices.
\begin{table}[h]
\caption{Relative timings for the different update schemes measured for simulations carried out on $16^4$ lattices. The significant computational overhead for the Metadynamics updates compared to the other algorithms is due to the stout smearing and the stout force recursion required for the Metadynamics force calculation.}
\label{tab:algorithm_benchmarks}
\begin{ruledtabular}
    \begin{tabular}{ccc}
        \multirow{2}{*}{Update scheme} & \multicolumn{2}{c}{Relative time} \\
                                       & Wilson action & DBW2 action       \\ \midrule
        1HB+4OR                        & 1             & 1                 \\
        1HMC                           & 2.90          & 4.21              \\
        MetaD-HMC (\texttt{4stout})    & 61.91         & 24.27             \\
        MetaD-HMC (\texttt{5stout})    & 73.64         & 28.11             \\
    \end{tabular}
\end{ruledtabular}
\end{table}
While no significant efforts were made to optimize the performance of our implementation of the MetaD-HMC, it is still clear that the additional overhead introduced by the computation of the Metadynamics force contribution is significant for pure gauge theory. The relative overhead is especially large compared to local update algorithms, which are already more efficient than the regular HMC. Note, however, that due to the more non-local character of the DBW2 gauge action, the relative loss in efficiency when switching to Metadynamics from either a local update algorithm or the HMC is already noticeably smaller.

Since the majority of the computational overhead comes from the Metadynamics force contribution, and the involved scales are different from those relevant for the gauge force, it seems natural to split the integration into multiple timescales in a similar fashion to the Sexton-Weingarten scheme \cite{Sexton:1992nu}: The force contributions from the bias potential are correlated to the topological charge, which is an IR observable, whereas the gauge force is usually dominated by short-range, UV fluctuations. Therefore, it is conceivable that integrating the Metadynamics force contribution on a coarser timescale than the gauge force could significantly decrease the required computational effort, while still being sufficiently accurate to lead to reasonable acceptance rates.

We have attempted to use combinations of both the Leapfrog and the Omelyan-Mryglod-Folk second-order integrator with the Omelyan-Mryglod-Folk fourth-order minimum norm integrator in a multiple timescale integration scheme. Unfortunately, we were unable to achieve a meaningful reduction of Metadynamics force evaluations without encountering integrator instabilities and deteriorating acceptance rates. However, this approach might still be helpful for simulations with dynamical fermions, where it is already common to split the forces into more than two levels.

Even if such a multiple timescale approach should turn out to be unsuccessful in reducing the number of Metadynamics force evaluations, we expect the relative overhead of Metadynamics to be much smaller for simulations including dynamical fermions. In a previous study \cite{Bonati:2017nhe} it was found that compared to conventional HMC simulations, simulations with Metadynamics using 20 steps of stout smearing were about three times slower in terms of real time.

\subsection{Scaling of the reweighting factor and improvements to the bias potential}
\label{subsec:5.2}
Due to the addition of the bias potential, unweighted averages do not lead to expectation values with respect to the original, physical probability density. If corrected with a reweighting procedure, the overlap between the sampled distribution and the distribution of physical interest needs to be sufficiently large for the method to work properly. A common measure to quantify the efficiency of the reweighting procedure is the effective sample size (ESS), defined as
\begin{equation}
    \mathrm{ESS} = \frac{\biggl(\sum\limits_i w_i \biggr)^2}{\sum\limits_i w_i^2},
    \label{eq:ESS}
\end{equation}
where $w_i$ is the respective weight associated with each individual configuration. In the case of a static bias potential, this is simply $e^{V(Q_{\mathrm{meta},i})}$. We found the normalized ESS, i.e., the ESS divided by the total number of configurations, to generally be of order $\mathcal{O}(10^{-2})$ or lower when simulating in regions of parameter space where conventional algorithms fail to sample different topological sectors.

Although the low ESS ultimately results from the fact that the bias potential is constructed in such a way as to have a flat marginal distribution over the CV, we can nonetheless distinguish between two contributions towards this effect. On the one hand, there is the inevitable flattening of the intersector barriers by the bias potential, which is necessary to facilitate tunneling between adjacent topological sectors. On the other hand, however, the different weights of the individual topological sectors are also canceled by the bias potential. While it is necessary to faithfully reproduce the intersector barriers, the leveling of the weights of the different topological sectors is often unwanted. It increases the time that the simulation spends at large values of $\abs{Q}$, so that these sectors are overrepresented compared to their true statistical weight. It is therefore conceivable that by retaining only the intersector barrier part of the bias potential, the relative weights of the different topological sectors will be closer to their physical values, and the ESS will increase. The resulting marginal distribution over the topological charge is then expected to no longer be constant, but rather resemble a parabola. In cases where this modification to the bias potential is used, we will either explicitly mention it or include the abbreviation \enquote{mod.} to make a clear distinction between it and the original potential.

Here and in \Cref{sec:6} of this work, we perform scaling tests of the proposed improvements in 2-dimensional U(1) gauge theory, where high statistics can be generated more easily than in 4-dimensional SU(3) gauge theory. The action is given by the standard Wilson plaquette action,
\begin{equation}
    S_g = \beta \sum\limits_{n \in \Lambda} \Bigl(1 - \Re\bigl[P_{t, x}(n)\bigr] \Bigr),
\end{equation}
and updates are performed with a single-hit Metropolis algorithm. The topological charge is defined using a geometric integer-valued definition:
\begin{equation}
    Q = \frac{1}{2\pi} \Im\Biggl[\sum\limits_{n \in \Lambda} \log P_{t, x}(n) \Biggr].
\end{equation}
For all Metadynamics updates, we use a field-theoretic definition of the topological charge that is generally not integer-valued:
\begin{equation}
    Q_{\mathrm{meta}} = \frac{1}{2\pi} \Im\Biggl[\sum\limits_{n \in \Lambda} P_{t, x}(n) \Biggr].
\end{equation}
Since the charge distributions obtained from the two definitions already show reasonable agreement without any smearing for the parameters considered here, we can use local update algorithms and directly include the Metadynamics contribution in the staple. A similar idea that encourages tunneling in the Schwinger model by adding a small modification to the action was proposed in \cite{deForcrand:1997fm}.

In previous tests in 2-dimensional U(1) gauge theory, we found that the bias potentials could be described by a sum of a quadratic and multiple oscillating terms \cite{Rouenhoff:2022seh}:
\begin{equation}
    V(Q) = A Q^2 + \sum\limits_{i = 1}^{N} B_{i} \sin^2(\pi f_{i} Q).
    \label{eq:parametric potential}
\end{equation}
Here, we fit our bias potentials that are obtained from the 2-dimensional U(1) simulations to this form. We then obtain a modified bias potential by subtracting the resulting quadratic term from the data.

\Cref{tab:U1scaling} contains the normalized ESS and integrated autocorrelation times for different lattice spacings on the same line of constant physics in 2-dimensional U(1) theory. We compare Metadynamics runs using bias potentials obtained directly from previous simulations with Metadynamics runs using potentials that were modified to retain the relative weights of the topological sectors as described above.
\begin{table}
\caption{Normalized effective sample sizes for simulations carried out on different lattices on the same line of constant physics in 2-dimensional U(1) gauge theory. For each set of parameters, \num{e7} measurements were performed with a separation of 10 update sweeps between every measurement. More details on the simulation setup can be found in \Cref{subsec:5.2}.}
\label{tab:U1scaling}
\begin{ruledtabular}
    \begin{tabular}{cccc}
        $L/a$ & $\beta$ & $\mathrm{ESS}/n_{\mathrm{meas}}$   & $\tau_{\mathrm{int}}(Q^2)$ \\ \toprule
        \multicolumn{4}{c}{Regular bias potential}                                        \\ \midrule
        16    & 3.2     & \SI{33.088 \pm 0.074}{\percent}    & 52                         \\
        20    & 5.0     & \SI{21.81 \pm 0.11}{\percent}      & 207                        \\
        24    & 7.2     & \SI{12.70 \pm 0.11}{\percent}      & 567                        \\
        28    & 9.8     & \SI{8.805 \pm 0.081}{\percent}     & 676                        \\
        32    & 12.8    & \SI{8.261\pm 0.085}{\percent}      & 1167                       \\ \toprule
        \multicolumn{4}{c}{Modified bias potential (see \Cref{subsec:5.2})}               \\ \midrule
        16    & 3.2     & \SI{99.50627\pm 0.00011}{\percent} & 5                          \\
        20    & 5.0     & \SI{64.3084\pm 0.0087}{\percent}   & 33                         \\
        24    & 7.2     & \SI{28.572\pm 0.012}{\percent}     & 123                        \\
        28    & 9.8     & \SI{27.291\pm 0.016}{\percent}     & 227                        \\
        32    & 12.8    & \SI{18.751\pm 0.025}{\percent}     & 247                        \\
    \end{tabular}
\end{ruledtabular}
\end{table}
We see large improvements for both the ESS and $\tau_{\mathrm{int}}$ in the modified case, even for the finest lattices considered.

We expect that the quadratic term is mostly relevant for small volumes and high temperatures. With larger volumes and lower temperatures, the slope should decrease, and with it the importance of correctly capturing this term. On the other hand, the oscillating term is expected to grow more important with finer lattice spacings, as the barriers between the different sectors grow steeper. Thus, the oscillating term needs to be described more and more accurately towards the continuum.

A standard technique to decrease, but not completely eliminate, action barriers is well-tempered Metadynamics \cite{PhysRevLett.100.020603}. In this approach, the height of the added Gaussians $w$ decays with increasing potential. In our tests, we found that this method increases the ESS at the cost of higher autocorrelation times. Whether the gains of the ESS outweigh the loss from the higher autocorrelation times depends on the choice of parameters. Although this technique might yield moderate improvements in the overall sampling efficiency, we decided not to attempt any fine-tuning of the parameters at this point.

\subsection{Accelerating the equilibration/buildup of the bias potential}
\label{subsec:5.3}
Another avenue of improvement is accelerating the buildup of the bias potential. This aspect becomes especially relevant when considering large-scale simulations, where current simulations are typically limited to $\mathcal{O}(10^4)$ update sweeps, and a lengthy buildup phase of the bias potential would render the method infeasible. Additionally, the range of relevant $Q$ values will also increase with larger physical volumes.

The first idea explored here is to exploit the aforementioned well-tempered variant of Metadynamics, by choosing a larger starting value of the Gaussian height $w$ and letting it decay slowly so as to minimize the change in the potential that arises from the decay. While this approach introduces the decay rate as another fine-tunable parameter, we found that this did indeed reduce the number of update iterations required to thermalize the potential. A small caveat is that an optimal choice of the decay rate requires prior knowledge on the approximate height of the action barriers.

\begin{figure}
    \centering
    \includegraphics{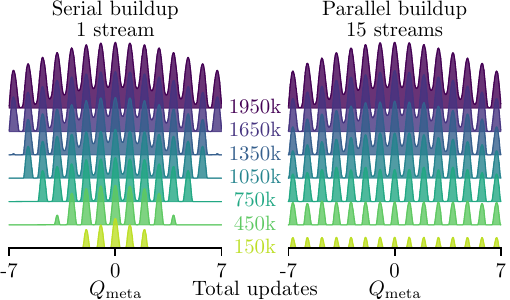}
    \caption{Comparison of serial and parallel buildup of the bias potential in 2-dimensional U(1) gauge theory for $32^2$ lattices. The ratio of update iterations per stream was held fixed at 15:1, so that both methods would use the same number of total update steps during each snapshot.}
    \label{fig:U1_parallel_build}
\end{figure}

A second way of reducing the buildup time is to use an enhancement of Metadynamics which is most commonly referred to as multiple walkers Metadynamics \cite{doi:10.1021/jp054359r}, where the potential is simultaneously built up by several independent streams in a trivially parallelizable way. In our case, it is convenient to make each stream start in a distinct topological sector. In 2-dimensional U(1) gauge theory, this can be achieved by seeding each stream with an instanton configuration of charge $Q$, which can be constructed according to \cite{Smit:1986fn}
\begin{equation}
    \begin{aligned}
        U_{t}^{I}(Q; t, x) &= \exp( -2 \pi i x \frac{Q}{N_x N_t}), \\
        U_{x}^{I}(Q; t, x) &= \exp(  2 \pi i t \frac{Q}{N_t} \delta_{x, N_x}).
    \end{aligned}
\end{equation}
The serial and parallel buildup are compared in \Cref{fig:U1_parallel_build} where the potential parameters for each stream are given by: $Q_{\mathrm{max/min}} = \pm7$, $n_{\mathrm{bins}} = 1400$, and $w = 0.002$.
In the case of 4-dimensional SU(3) the direct construction of instantons with higher charge is not quite as simple as in 2-dimensional U(1) gauge theory. The construction of lattice instantons with even charge is described in \cite{Smit:1986fn}, and lattice instantons with odd charge can be constructed by combining multiple instantons with charge $\abs{Q} = 1$ \cite{Jahn:2019nmd, Eichhorn:2022wxn}. Regardless, starting with instantons is not required, since we only need each stream to fall into the specified sector. The time until the streams start to tunnel is a first indicator of the thermalization timescale of the potential.

Independent of the possible improvements mentioned here, a fine-tuning of the standard Metadynamics parameters could also prove to be worthwhile in regard to accelerating the buildup and improving the quality of the bias potential.
\section{Combining Metadynamics with parallel tempering}
\label{sec:6}
\begin{figure}[h]
    \centering
    \vspace{-0.5cm}
    \input{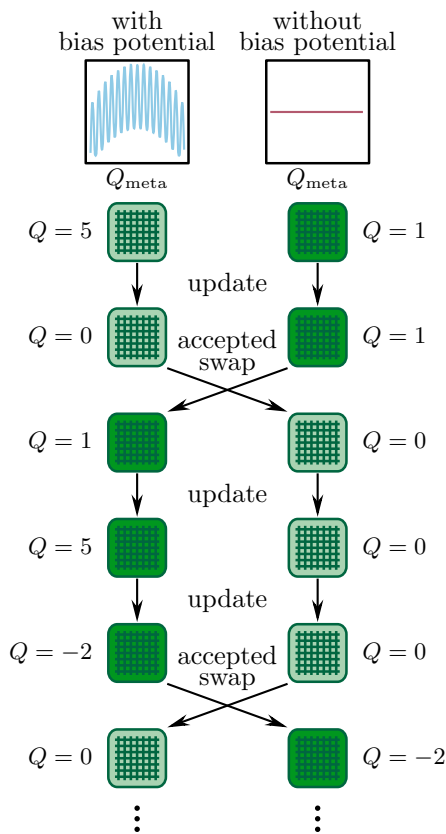}
    \caption{Illustration of the PT-MetaD algorithm, with the two columns representing two simulation streams. The bias potentials are illustrated at the top, while subsequent rows represent the time series, with different shades of the configurations for better visibility. $Q$ values are indicative and do not represent results from an actual simulation.}
    \label{fig:PT-MetaD_illustration}
\end{figure}
In order to eliminate the problem of small effective sample sizes observed in our Metadynamics simulations due to the required reweighting, we propose to combine Metadynamics with parallel tempering \cite{PhysRevLett.57.2607}. This is done in a spirit similar to the parallel tempering on a line defect proposed by Hasenbusch \cite{Hasenbusch:2017unr}. We introduce two simulation streams: One with a bias potential, and the other without it, while the physical actions $S(U)$ are the same for both streams, as illustrated in \Cref{fig:PT-MetaD_illustration}. Since we are working in pure gauge theory, this means the second stream without bias potential can be simulated with local update algorithms. After a fixed number of updates have been performed on the two streams, a swap of the configurations is proposed and subject to a standard Metropolis accept-reject step. The action difference is given by
\begin{equation}
    \begin{aligned}
        \Delta S^M_t &= \bigl[S^M_{t}(U_1) + S(U_2)\bigr] - \bigl[S^M_{t}(U_2) + S(U_1)\bigr]
        \\&= V_t(Q_{\mathrm{meta}, 1}) - V_t(Q_{\mathrm{meta}, 2}),
        \label{eq:pt_action_diff}
    \end{aligned}
\end{equation}
where the indices of the quantities denote the number of the stream, and $V_t$ is the bias potential in the first stream. It is apparent and important to note that the action difference is simple to compute regardless of what the physical action looks like. Even in simulations where dynamical fermions are present, the contributions from the physical action are always canceled out by virtue of the two streams having the same action parameters; only the contribution from the bias potential remains. Moreover, the action differences, and thus the swap rates, should be largely independent of the volume.

Since the second/measurement stream samples configurations according to the physical distribution no reweighting is needed, and thus the effective sample size is not reduced. Additionally, if the swaps are effective, the measurement stream will inherit the topological sampling from the stream with bias potential and thus also sample topological sectors well. Effectively, the accept-reject step for swap proposals serves as a filter for configurations with vanishing weight, thereby decreasing the statistical uncertainties on all observables weakly correlated to the topological charge. What remains to be seen is whether the efficiency of the sampling of the topological sectors carries over from the bias potential stream to the measurement stream. In this section, we address this question via both scaling tests in 2-dimensional U(1) and exploratory runs in 4-dimensional SU(3) in a region where conventional update algorithms are effectively frozen.

\subsection{Scaling tests in 2-dimensional U(1)}
\label{subsec:6.1}
We carried out a number of simulations in 2-dimensional U(1) gauge theory using the same parameters and simulation setup as described in \Cref{subsec:5.2}. We use bias potentials already built for these Metadynamics runs and keep them static in a number of parallel tempered Metadynamics runs. For each set of parameters, we carried out one run with the respective unmodified potential and one run with a potential modified as described in \Cref{subsec:5.2}. In these runs, swaps between the two streams were proposed after each had completed a single update sweep over all lattice sites. The run parameters, as well as the resulting autocorrelation times of the topological charge $Q$, can be found in \Cref{tab:U1PTscaling}.

Since the relevant configuration space is now the product of configuration spaces of the individual streams, autocorrelation timescales of observables that are defined on the product space will now contain important information about the dynamics of the system. Therefore, we additionally monitor the sum of the squared topological charges on both streams. This observable allows us to distinguish the fluctuations in $Q$ originating from true tunneling events from repeated swaps between the two streams without tunneling.
\begin{table}
\caption{Integrated autocorrelation times for different lattices on the same line of constant physics in 2-dimensional U(1) gauge theory, using both Metropolis and PT-MetaD updates. Observables indexed with $1$ are taken from the stream with bias potential, whereas those indexed with $2$ are taken from the regular stream. The modification of the bias potential is discussed in \Cref{subsec:5.2}. Overall, $10^7$ measurements were performed with a separation of 10 update sweeps between every measurement.}
\label{tab:U1PTscaling}
\begin{ruledtabular}
    \begin{tabular}{cccc}
        $L/a$ & $\beta$ & $\tau_{\mathrm{int}}(Q_2^2)$ & $\tau_{\mathrm{int}}(Q_1^2 + Q_2^2)$ \\
        \toprule
        \multicolumn{4}{c}{Single-hit Metropolis}                                           \\ \midrule
        16  & 3.2     & 3                            & N/A                                  \\
        20  & 5.0     & 71                           & N/A                                  \\
        24  & 7.2     & 3939                         & N/A                                  \\
        28  & 9.8     & 462472                       & N/A                                  \\
        32  & 12.8    & $>$10000000                      & N/A                                  \\
        \toprule
        \multicolumn{4}{c}{PT-MetaD (regular bias potential)}                               \\ \midrule
        16  & 3.2     & 5                            & 20                                   \\
        20  & 5.0     & 59                           & 142                                  \\
        24  & 7.2     & 939                          & 534                                  \\
        28  & 9.8     & 1730                         & 818                                  \\
        32  & 12.8    & 1926                         & 1319                                 \\
        \toprule
        \multicolumn{4}{c}{PT-MetaD (modified bias potential)}                              \\ \midrule
        16  & 3.2     & 4                            & 5                                    \\
        20  & 5.0     & 47                           & 47                                   \\
        24  & 7.2     & 184                          & 208                                  \\
        28  & 9.8     & 316                          & 403                                  \\
        32  & 12.8    & 312                          & 465                                  \\
    \end{tabular}
\end{ruledtabular}
\end{table}

\begin{figure}
    \centering
    \includegraphics{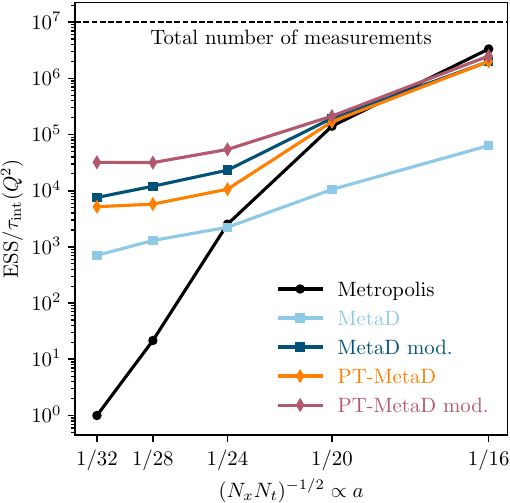}
    \caption{Continuum scaling of the total sample size for the standard Metropolis algorithm and variations of MetaD-Metropolis in 2-dimensional U(1), including the use of a modified bias potential (see \Cref{subsec:5.2}). The corresponding data can be found in \Cref{tab:U1scaling} and \Cref{tab:U1PTscaling}. Lines are drawn to guide the eyes.}
    \label{fig:U1_scaling}
\end{figure}

\Cref{fig:U1_scaling} shows the scaling of the total amount of independent configurations, which is given by the quotient of the effective sample size \Cref{eq:ESS} and the integrated autocorrelation time of the topological susceptibility. The performance of the standard Metropolis algorithm is compared to parallel tempered and standard Metadynamics, with both modified (see \Cref{subsec:5.2}) and non-modified bias potentials.

PT-MetaD performs well throughout the entire range of lattice spacings, consistently outperforming standard Metadynamics by more than an order of magnitude. Most importantly, the ratio of independent configurations seems to reach a plateau for finer lattice spacings, indicating an improved scaling behavior compared to conventional Metadynamics, which in itself already scales significantly better than the single-hit Metropolis algorithm. It is also worth noting that the modified bias potential provides better results than the non-modified one, as evidenced by the normalized effective sample sizes and swap rates presented in \Cref{tab:U1_swaprates}. This is consistent with our expectation that large excursions in the topological charge, which produce irrelevant configurations, are curbed by the modification of the bias potential.
\begin{table}
\caption{Normalized effective sample sizes and swap rates for PT-MetaD simulations with regular and modified bias potential in 2-dimensional U(1) gauge theory, where the normalized effective sample sizes were measured on the bias potential streams of the PT-MetaD runs. The non-monotonous behaviour of the quantities for the PT-MetaD runs may be due to a suboptimal choice of bias potentials.}
\label{tab:U1_swaprates}
\begin{ruledtabular}
    \begin{tabular}{cccccc}
            &         & \multicolumn{2}{c}{PT-MetaD}                 & \multicolumn{2}{c}{PT-MetaD mod.}            \\
        \cmidrule(l){3-4} \cmidrule(l){5-6}
        $L/a$ & $\beta$ & $\mathrm{ESS}/n_{\mathrm{meas}}$ & Swap rate & $\mathrm{ESS}/n_{\mathrm{meas}}$ & Swap rate \\
        \midrule
        16  & 3.2     & \SI{21.0}{\percent}              & 0.198     & \SI{99.6}{\percent}              & 0.960     \\
        20  & 5.0     & \SI{13.6}{\percent}              & 0.151     & \SI{64.3}{\percent}              & 0.585     \\
        24  & 7.2     & \SI{7.3}{\percent}               & 0.086     & \SI{28.6}{\percent}              & 0.271     \\
        28  & 9.8     & \SI{5.0}{\percent}               & 0.057     & \SI{27.3}{\percent}              & 0.249     \\
        32  & 12.8    & \SI{8.3}{\percent}               & 0.081     & \SI{18.8}{\percent}              & 0.169
    \end{tabular}
\end{ruledtabular}
\end{table}
\begin{figure*}
    \centering
    \includegraphics{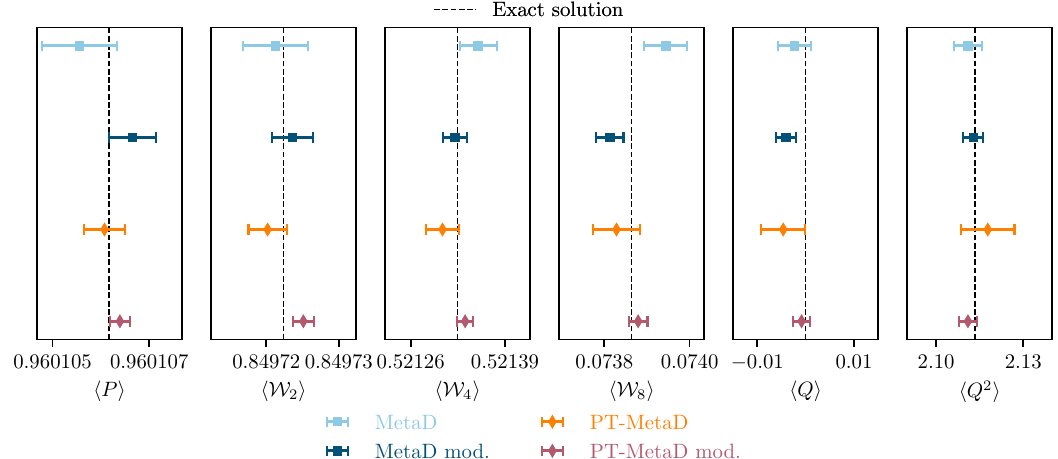}
    \caption{Comparison of expectation values and uncertainties for the plaquette $P$, larger Wilson loops $\mathcal{W}_{n}$, the topological charge $Q$, and the topological susceptibility $Q^2$ for variations of MetaD-Metropolis in 2-dimensional U(1) for $32^2$ lattices and $\beta = 12.8$. The modification of the bias potential is discussed in \Cref{subsec:5.2}. Dashed lines correspond to the exact solutions \cite{Kovacs:1995nn, Elser:2001pe, Bonati:2019ylr, Albandea:2021lvl}. Results obtained from a run with standard Metropolis updates are off the scale everywhere except for $Q$, where the expectation value is exactly equal to zero due to complete topological freezing.}
    \label{fig:U1_observables}
\end{figure*}
A more detailed look at the effectiveness of the new algorithm is provided by \Cref{fig:U1_observables}. It compares the results of PT-MetaD and standard Metadynamics at our finest lattice spacing, with and without modification of the bias potential. For reference, exact values from analytical solutions \cite{Kovacs:1995nn, Elser:2001pe, Bonati:2019ylr, Albandea:2021lvl} are also provided (see \Cref{appendix:U1_exact_solution} for more details). First, we note that there is no significant difference in the performance between standard and parallel tempered Metadynamics in the topology related observables $Q$ and $Q^2$ in the case of a modified bias potential. This indicates that the swaps are effective in carrying over the topological sampling of the bias potential stream to the measurement stream. On the other hand, the inclusion of the irrelevant higher sectors with the unmodified bias potential does increase the error bars, and there is some indication that not all of the topological sector sampling is carried over into the measurement run of PT-MetaD. Furthermore, \Cref{fig:U1_observables} reveals that for observables not related to the topology, PT-MetaD with a modified bias potential is superior to standard Metadynamics. This is clearly the effect of the higher effective sample size and number of independent configurations.

In summary, our scaling tests in 2-dimensional U(1) suggest that parallel tempered Metadynamics with a modified bias potential has much improved topological sampling, equivalent to standard Metadynamics, while at the same time not suffering from a reduced effective sample size. There is some indication that the ratio of statistically independent to total configurations does reach a stable plateau in the continuum limit.

\subsection{First results in 4-dimensional SU(3)}
\label{subsec:6.2}
\begin{figure*}
    \centering
    \includegraphics{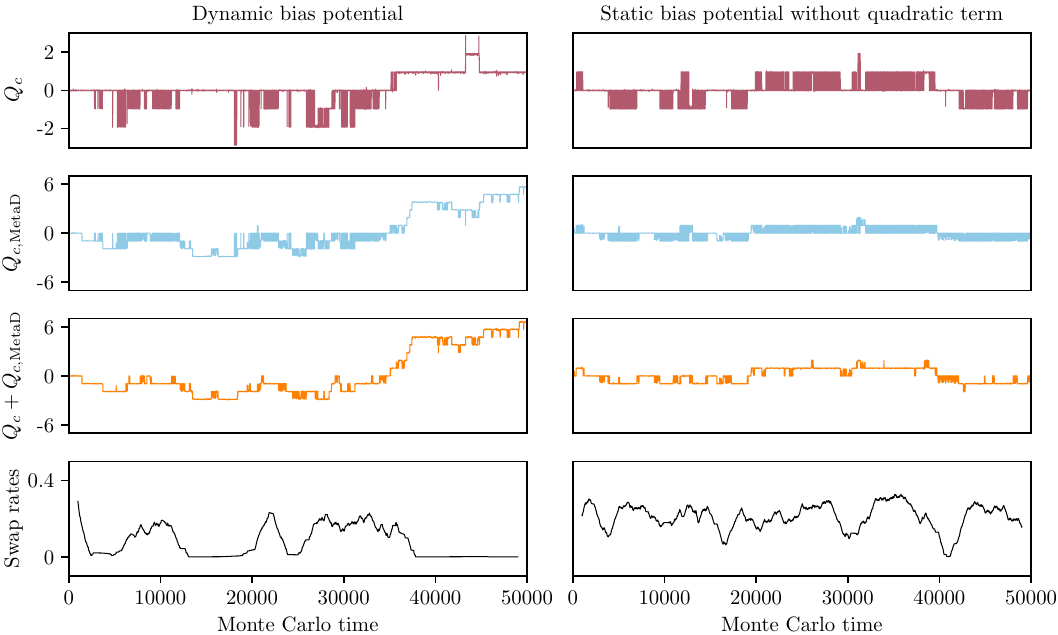}
    \caption{Topological charge time series for our parallel tempered Metadynamics runs on a $V = 16^4$ lattice at $\beta = 1.25$ with the DBW2 gauge action and four steps of stout smearing in the definition of $Q_\mathrm{meta}$. The left panels show results of our first run with a dynamically built bias potential, while the right panels show our second run with a modified static potential. The topmost row shows the time series of the topological charges in the respective measurement runs, while the second row is for the Metadynamics part (note the different $y$-scales). The third row displays the sum of the topological charges on the measurement stream and the stream including a bias potential, serving as an indicator for genuine transitions of the entire system into new topological sectors. In the bottom row, the running average of the swap acceptance rate with a window size of 2000 is displayed.}
    \label{fig:tempering_timeseries}
\end{figure*}

For our exploratory study in 4-dimensional SU(3), we turn to the DBW2 gauge action at $\beta = 1.25$ on a $V = 16^4$ lattice, which we have already used in \Cref{sec:5}. For our first run, which is depicted in the left panels of \Cref{fig:tempering_timeseries}, we have combined a local 1HB+4OR measurement stream with a \texttt{4stout} MetaD-HMC stream that dynamically generates the bias potential. Between swap proposals, updates for the two streams are performed at a ratio of 10 1HB+4OR update sweeps to a single unit length MetaD-HMC trajectory, which roughly reflects the relative wall clock times between the algorithms. One can see that the measurement run starts exploring other topological sectors almost as soon as the parallel run with active bias potential has gained access to them.
\begin{figure}
    \centering
    \includegraphics{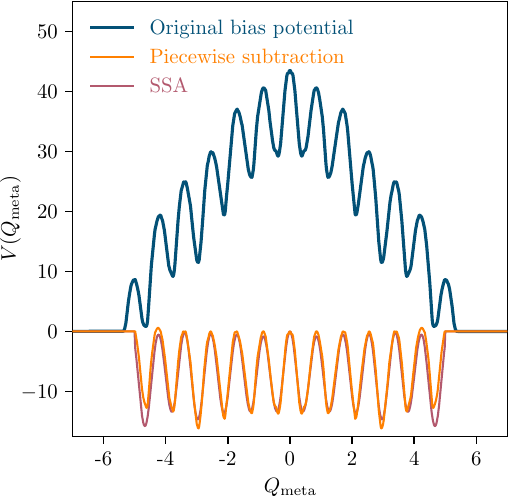}
    \caption{Comparison between the original bias potential and its trend subtracted modifications from singular spectrum analysis and piecewise subtraction of the $Q^2$ term.}
    \label{fig:bias_potential_extraction}
\end{figure}
In the later stages of the run, when the bias potential is sufficiently built up to allow the Metadynamics run to enter higher topological sectors, one can see that the swap rate is lowered by the action difference between the topological sectors, leading to an overall swap rate of $\sim \SI{8}{\percent}$. This effect mirrors the reduction of the effective sample size in pure Metadynamics updates and may be ameliorated by removing the quadratic term in the bias potential, as discussed in \Cref{subsec:5.2}. In fact, the relevant point is that the action difference between the maxima of the bias potential for different topological sectors reflects the relative weight of these sectors in the path integral and should not be flattened out. Ideally, we want the bias potential to only reproduce the barriers between the sectors, not their relative weights. For a second exploratory PT-MetaD run, we therefore opted for a static bias potential of this sort. Lacking data that are precise enough to model the bias potential in detail, as we did in 2-dimensional U(1), we started from the bias potential of a previous Metadynamics run and extracted the high frequency (in $Q_\mathrm{meta}$) part corresponding to the topological barriers, while eliminating the long range part corresponding to the relative weights of the topological sectors. For this purpose, we chose to perform a singular spectrum analysis \cite{VAUTARD1989395} and crosschecked the result with a simple piece-wise subtraction of the $Q^2$ term between consecutive local maxima. As displayed in \Cref{fig:bias_potential_extraction}, both methods result in a similar modified bias potential that seems to reproduce the intersector barriers rather well.
The right panels of \Cref{fig:tempering_timeseries} display the results of the corresponding PT-MetaD run. Notably, excursions to large absolute values of the topological charge in the stream with bias potential are now curbed, and the swap acceptance rate has increased to $\sim \SI{21}{\percent}$. In addition, the acceptance rate is approximately constant over the entire run, as it should be expected for a static bias potential. The resulting autocorrelation times are $\tau_{\mathrm{int}}(Q_c^2) = \num{53 \pm 3}$ and $\tau_{\mathrm{int}}(Q_c^2 + Q_{c, \textrm{MetaD}}^2) = \num{333 \pm 35}$. We would like to emphasize that the bias potential we extracted is a rather rough guess. With a larger amount of data, it might be possible to extract a better bias potential, possibly leading to even higher acceptance rates. Considering the rather simple forms used to model the bias potential, it might also be possible to describe it with sufficient accuracy for good initial guesses at other run parameters. We plan to address these points in the future.

In any case, these first results clearly show that the parallel tempered Metadynamics algorithm is able to achieve enhanced topological sampling in 4-dimensional SU(3) without the reduction of the effective sample size that is typical for algorithms with a bias potential.
\newpage
\section{Conclusion and outlook}
\label{sec:7}
In this paper, we have proposed a new update algorithm, parallel tempered Metadynamics, PT-MetaD and applied it to 4-dimensional SU(3) gauge theory. In its simplest form, which we have investigated here, it consists of two parallel streams simulating the same physical system. One of the streams includes a fixed bias potential, which facilitates transitions between topological sectors. This bias potential can, for example, be extracted from a Metadynamics simulation of the system. The second stream can utilize any efficient conventional update algorithm that, by itself, may exhibit topological freezing. At regular intervals, swaps between the two streams are proposed and subjected to a Metropolis accept-reject step. We have demonstrated that in this way, the good topological sampling of the stream with bias potential carries over to the conventional update stream without compromising its effective sample size. When performing measurements exclusively on the second, conventional update stream, one therefore obtains topological unfreezing without any reweighting or additional modifications. The price to pay is the additional update stream including a bias potential, which requires minimal communication with the measurement stream and thus is embarrassingly parallel. The proposed algorithm may be helpful in overcoming potential barriers in more general cases without compromising the effective sample size.

We have demonstrated that PT-MetaD can unfreeze 4-dimensional SU(3) gauge theory at parameter values where conventional algorithms are frozen. Furthermore, scaling tests in 2-dimensional U(1) gauge theory indicate gains of more than an order of magnitude compared to standard Metadynamics in the total sample size and an improved scaling of autocorrelation times with the lattice spacing compared to standard update algorithms. We have also demonstrated that the buildup of the Metadynamics bias potential may be accelerated by running multiple Metadynamics simulations in parallel.

We believe these results are promising and plan to study the scaling behavior of the methods tested here in more detail for 4-dimensional SU(3) gauge theory, and eventually in full QCD. Conceptually, there seem to be no obstacles for implementing parallel tempered Metadynamics in full QCD. We also plan to explore possible optimizations for parallel tempered Metadynamics. These include optimizing the bias potential via enhanced buildup and extraction and, possibly, describing it parametrically. Furthermore, it would be interesting to investigate whether adding additional streams with the same or modified bias potentials could increase performance, despite the additional computational cost. This might be especially interesting for large scale simulations, where additional streams may offer a reduction of autocorrelation times in a trivially parallelizable manner.
\newpage

\begin{acknowledgments}
We thank Philip Rouenhoff for collaboration in the early stages of this work. We gratefully acknowledge helpful discussions with Szabolcs Borsanyi, Stephan Dürr, Fabian Frech, Jana Günther, Ruben Kara, Andrey Kotov, and Kalman Szabo. Calculations were performed on a local PC cluster at the University of Wuppertal.
\end{acknowledgments}

\appendix
\label{appendix}
\section{Conventions}
\label{appendix:Conventions}
We consider the Lie group SU(3) along with the associated Lie algebra $\mathfrak{su}(3)$. The generators $T^a$ may be represented as traceless anti-Hermitian $3 \times 3$ matrices with the group index $a \in \{1, \dots, 8\}$. In terms of the Gell-Mann matrices, the generators may be chosen to be
\begin{equation}
    T^a = \frac{i}{2} \lambda^a,
\end{equation}
but more generally, the generators are normalized in such a way that
\begin{equation}
    \tr[T^a T^b] = -\frac{1}{2} \delta_{ab}.
\end{equation}
The scalar product on the algebra is defined as
\begin{equation}
    \langle A, B \rangle = - 2 \tr[A B] = A^{a} B^{a},
\end{equation}
where summation over the group indices $a$ is implied. Note that the definition of the scalar product affects the trajectory length in the HMC and the MetaD-HMC. The convention we use throughout this paper agrees with, e.g., \cite{Schaefer:2010hu}.

We also introduce the following abbreviation for the projector that maps any $3 \times 3$ matrix onto $\mathfrak{su}(3)$:
\begin{equation}
    \begin{aligned}
        \mathcal{P}_{\mathfrak{su}(3)}(A) &= \frac{1}{2}(A - A^{\dagger}) - \frac{1}{6}\tr[A - A^{\dagger}] \\
                                          &= -2 T^a \Re\tr[T^a A].
        \label{eq:su3_algebra_projector}
    \end{aligned}
\end{equation}
The derivative of a scalar function of the gauge field $f(U)$ with respect to a single link $U_{\alpha}(n)$ is defined as a linear combination of generators $T^a$ and partial derivative operators $\partial_{n, \alpha}^{a}$:
\begin{equation}
    \partial_{n, \alpha} f(U) = T^{a} \partial_{n, \alpha}^{a} f(U).
    \label{eq:Link_derivative}
\end{equation}
Individually, the action of the $a$th partial derivative operator on $f(U)$ is defined as
\begin{equation}
    \partial_{n, \alpha}^{a} f(U) = \frac{\mathrm{d}}{\mathrm{d}s} f(U^{[s]}) \bigg\vert_{s = 0},
\end{equation}
where
\begin{equation}
    U_{\nu}^{[s]}(m) =
    \begin{cases}
		e^{s T^{a}} U_{\alpha}(n) & \text{if } (m, \nu) = (n, \alpha),\\
		U_{\nu}(m)             & \text{else.}
    \end{cases}
\end{equation}
While the result of the partial derivatives acting on a function is basis-dependent, the linear combination in \Cref{eq:Link_derivative} is not.

\section{Clover charge derivative}
\label{appendix:Clover_charge_derivative}
In order to obtain an expression for the force contribution from the topological bias potential in \Cref{eq:metadynamics_force}, the algebra-valued derivative of the collective variable $Q_{\mathrm{meta}}$ with respect to the group-valued fully smeared links has to be calculated first. Evidently, this part of the force calculation does not depend on the smearing procedure used. The subsequent stout force recursion \cite{Morningstar:2003gk, BMW:2010skj} required to relate this term to the derivative with respect to unsmeared links is discussed in \Cref{appendix:Stout_force_recursion}.

Recall that the clover-based definition of the topological charge is given by
\begin{equation}
    Q_c = \frac{1}{32 \pi^2} \sum_{n \in \Lambda} \epsilon_{\mu \nu \rho \sigma} \tr[\Hat{F}^{\mathrm{clov}}_{\mu \nu}(n) \Hat{F}^{\mathrm{clov}}_{\rho \sigma}(n)],
    \label{eq:top_charge_clov_def_appendix}
\end{equation}
where the field strength tensor is defined as
\begin{equation}
    \Hat{F}^{\mathrm{clov}}_{\mu \nu}(n) = -\frac{i}{8} \bigl(C_{\mu \nu}(n) - C_{\nu \mu}(n) \bigr),
    \label{eq:F_clov_def_appendix}
\end{equation}
and the clover term is given by
\begin{equation}
    C_{\mu \nu}(n) = P_{\mu, \nu}(n) + P_{\nu, -\mu}(n) + P_{-\mu, -\nu}(n) + P_{-\nu, \mu}(n).
\end{equation}
For notational convenience, we introduce the auxiliary variables $R_{\mu \nu}(n) = C_{\mu \nu}(n) - C_{\nu \mu}(n)$ and drop the specification of the lattice site $n$ unless pertinent to the formula.

What we need for the force is the sum of the partial derivatives in all eight group directions. By using the cyclicity of the trace and the symmetry relation
\begin{equation}
    R_{\mu \nu}^{\dagger} = -R_{\mu \nu} = R_{\nu \mu},
\end{equation}
we can rewrite the sum of partial derivatives as
\begin{equation}
    \begin{aligned}
        &T^a \partial_{n, \alpha}^{a} \epsilon_{\mu \nu \rho \sigma} \tr[R_{\mu \nu} R_{\rho \sigma}]
        \\= 4 &T^a \sum\limits_{\nu \rho \sigma} \partial_{n, \alpha}^{a} \epsilon_{\alpha \nu \rho \sigma} \tr[R_{\alpha \nu} R_{\rho \sigma}].
    \end{aligned}
    \label{eq:clov_partial_derivative_sum}
\end{equation}
Note that in the second line of \Cref{eq:clov_partial_derivative_sum} and the subsequent equations in this section, the index $\alpha$ is not summed over.

Applying the partial derivative operators, we get two nonvanishing contributions from every plaquette term. Using the same symmetry arguments as before, we can further simplify the expression to obtain the following result for the derivative:
\begin{widetext}
\begin{equation}
	\begin{aligned}
		  4 T^a \sum\limits_{\nu \rho \sigma} \partial_{n, \alpha}^{a} \epsilon_{\alpha \nu \rho \sigma} \tr[R_{\alpha \nu} R_{\rho \sigma}]
              =& \; 8 T^a \sum\limits_{\nu \rho \sigma} \epsilon_{\alpha \nu \rho \sigma} \Re \tr\Bigl[ T^{a} U_{\alpha}(n) U_{\nu}(n + \alpha) U^{\dagger}_{\alpha}(n + \nu) U^{\dagger}_{\nu}(n) R_{\rho \sigma}(n)
		\\ &- T^{a} U_{\alpha}(n) U^{\dagger}_{\nu}(n + \alpha - \nu) U^{\dagger}_{\alpha}(n - \nu) R_{\rho \sigma}(n - \nu) U_{\nu}(n - \nu)
		\\ &- T^{a} U_{\alpha}(n) U^{\dagger}_{\nu}(n + \alpha - \nu) R_{\rho \sigma}(n + \alpha - \nu) U^{\dagger}_{\alpha}(n - \nu) U_{\nu}(n - \nu)
		\\ &+ T^{a} U_{\alpha}(n) R_{\rho \sigma}(n + \alpha) U_{\nu}(n + \alpha) U^{\dagger}_{\alpha}(n + \nu) U^{\dagger}_{\nu}(n)
		\\ &- T^{a} U_{\alpha}(n) U^{\dagger}_{\nu}(n + \alpha - \nu) U^{\dagger}_{\alpha}(n - \nu) U_{\nu}(n - \nu) R_{\rho \sigma}(n)
		\\ &+ T^{a} U_{\alpha}(n) U_{\nu}(n + \alpha) U^{\dagger}_{\alpha}(n + \nu) R_{\rho \sigma}(n + \nu) U^{\dagger}_{\nu}(n)
		\\ &- T^{a} U_{\alpha}(n) R_{\rho \sigma}(n + \alpha) U^{\dagger}_{\nu}(n + \alpha - \nu) U^{\dagger}_{\alpha}(n - \nu) U_{\nu}(n - \nu)
		\\ &+ T^{a} U_{\alpha}(n) U_{\nu}(n + \alpha) R_{\rho \sigma}(n + \alpha + \nu) U^{\dagger}_{\alpha}(n + \nu) U^{\dagger}_{\nu}(n) \Bigr]
            \\=& \; 8 T^a \sum\limits_{\nu \rho \sigma} \epsilon_{\alpha \nu \rho \sigma} \Re\tr[T^{a} A_{\alpha\nu\rho\sigma}]
            \\=& \; 8 T^a \Re\tr[T^{a} A_{\alpha}].
	\end{aligned}
\end{equation}
\end{widetext}
An expression of the above form can be rewritten using the projector induced by the scalar product of the algebra defined in \Cref{eq:su3_algebra_projector}:
\begin{equation}
    8 T^a \Re\tr[T^{a} A_{\alpha}] = -4 \mathcal{P}_{\mathfrak{su}(3)}(A_{\alpha}).
\end{equation}
Including the missing prefactors from \Cref{eq:top_charge_clov_def_appendix} and \Cref{eq:F_clov_def_appendix}, we obtain the following final result for the derivative of the clover-based topological charge with respect to the gauge link $U_{\alpha}(n)$:
\begin{equation}
    \begin{aligned}
        T^a \partial_{n, \alpha}^{a} Q_c &= \frac{1}{32\pi^2} T^a \partial_{n, \alpha}^{a} \epsilon_{\mu \nu \rho \sigma} \tr[\Hat{F}^{\mathrm{clov}}_{\mu \nu} \Hat{F}^{\mathrm{clov}}_{\rho \sigma}] \\
        &= -\frac{1}{2048\pi^2} T^a \partial_{n, \alpha}^{a} \epsilon_{\mu \nu \rho \sigma} \tr[R_{\mu \nu} R_{\rho \sigma}] \\
        &= \frac{1}{512\pi^2} \mathcal{P}_{\mathfrak{su}(3)}(A_{\alpha}).
    \end{aligned}
\end{equation}
In a similar way, expressions for the derivatives of improved definitions of the topological charge involving larger loops can be obtained.

\section{Stout force recursion}
\label{appendix:Stout_force_recursion}
The derivative of the topological charge with respect to an unsmeared link can be computed by propagating the initial derivative with respect to a fully smeared link through the different smearing levels \cite{Morningstar:2003gk, BMW:2010skj}. This part of the force calculation is completely independent of the topological charge operator and only depends on the smearing procedure used. In fact, the recursion is exactly the same as the recursion used during the force calculation with stout smeared fermions. Note that starting the calculation at the highest smearing level is vastly more efficient, despite having to track intermediate smearing levels, since the function's domain is high-dimensional, whereas the codomain is 1-dimensional.

We begin by establishing the required definitions before describing the actual force recursion. For notational convenience, we only consider a single smearing step without loss of generality and denote smeared quantities with a prime symbol. For the most part, we will use the same notation as in \cite{Morningstar:2003gk}. Starting from a gauge field $U$, the stout smeared gauge field $U'$ is defined as
\begin{align}
    U'_{\mu}(n)     &= e^{i Q_{\mu}(n)} U_{\mu}(n),                                  \\
    Q_{\mu}(n)      &= -i \mathcal{P}_{\mathfrak{su}(3)}\bigl(\Omega_{\mu}(n)\bigr), \\
    \Omega_{\mu}(n) &= C_{\mu}(n) U_{\mu}^{\dagger}(n).
    \label{eq:Omega_stout}
\end{align}
Note that no summation is performed over $\mu$ in \Cref{eq:Omega_stout}. $C_{\mu}(n)$ is a weighted sum of staples:
\begin{equation}
    \begin{aligned}
        C_{\mu}(n) = \sum\limits_{\nu \neq \mu} \rho_{\mu \nu} &\Bigl( U_{\nu}(n) U_{\mu}(n + \Hat{\nu}) U_{\nu}^{\dagger}(n + \Hat{\mu}) \\[-2ex]
                                                               &+ U_{\nu}^{\dagger}(n - \Hat{\nu}) U_{\mu}(n - \Hat{\nu}) U_{\nu}(n + \Hat{\mu} - \Hat{\nu}) \Bigr).
    \end{aligned}
\end{equation}
If the smeared force $F'$ is known, the unsmeared force $F$ is given by
\begin{equation}
    \begin{aligned}
        F_\mu(n) =& \; \mathcal{P}_{\mathfrak{su}(3)}\biggl(U_{\mu}(n) \Sigma'_{\mu}(n) e^{i Q_{\mu}(n)}
        \\&+ i U_{\mu}(n) C_{\mu}^{\dagger}(n) \Lambda_{\mu}(n) - i U_{\mu}(n) \Upsilon_{\mu}(n) \biggr),
    \end{aligned}
    \label{eq:stout_force_recursion}
\end{equation}
where
\begin{equation}
    \begin{aligned}
        \Upsilon_{\mu}(n) &= \sum\limits_{\nu \neq \mu} \Bigl( \rho_{\nu \mu}U_{\nu}(n + \Hat{\mu}) U_{\mu}^{\dagger}(n + \Hat{\nu}) U_{\nu}^{\dagger}(n) \Lambda_{\nu}(n)
        \\  +& \rho_{\mu \nu}U_{\nu}^{\dagger}(n + \Hat{\mu} - \Hat{\nu}) U_{\mu}^{\dagger}(n - \Hat{\nu}) \Lambda_{\mu}(n - \Hat{\nu}) U_{\nu}(n - \Hat{\nu})
        \\  +& \rho_{\nu \mu}U_{\nu}^{\dagger}(n + \Hat{\mu} - \Hat{\nu}) \Lambda_{\nu}(n + \Hat{\mu} - \Hat{\nu}) U_{\mu}^{\dagger}(n - \Hat{\nu}) U_{\nu}(n - \Hat{\nu})
        \\  -& \rho_{\nu \mu}U_{\nu}^{\dagger}(n + \Hat{\mu} - \Hat{\nu}) U_{\mu}^{\dagger}(n - \Hat{\nu}) \Lambda_{\nu}(n - \Hat{\nu}) U_{\nu}(n - \Hat{\nu})
        \\  -& \rho_{\nu \mu}\Lambda_{\nu}(n + \Hat{\mu}) U_{\nu}(n + \Hat{\mu}) U_{\mu}^{\dagger}(n + \Hat{\nu}) U_{\nu}^{\dagger}(n)
        \\  +& \rho_{\mu \nu}U_{\nu}(n + \Hat{\mu}) U_{\mu}^{\dagger}(n + \Hat{\nu}) \Lambda_{\mu}(n + \Hat{\nu}) U_{\nu}^{\dagger}(n) \Bigr),
    \end{aligned}
\end{equation}
\begin{align}
    \Lambda =& \; \frac{1}{2}(\Gamma + \Gamma^{\dagger}) - \frac{1}{6} \tr[\Gamma + \Gamma^{\dagger}], \label{eq:stout_force_lambda} \\
    \begin{split}
    \Gamma  =& \; \tr[B_1 U \Sigma' ] Q + \tr[B_2 U \Sigma'] Q^2                                                                     \\
             &+ f_1 U \Sigma' + f_2 Q U \Sigma' + f_2 U \Sigma' Q,
    \end{split} \label{eq:stout_force_gamma}                                                                                         \\
    \Sigma' =& \; U^{\prime\, \dagger} F. \label{eq:stout_force_sigma}
\end{align}
The exact form of the $f_j$ and $B_i$ is given in \Cref{appendix:Cayley-Hamilton_exponential}. For multiple smearing levels, the procedure in \Cref{eq:stout_force_recursion} can simply be repeated.

\section{Computation of the matrix exponential using the Cayley-Hamilton theorem}
\label{appendix:Cayley-Hamilton_exponential}
The computation of the matrix exponential is done along the lines of \cite{Morningstar:2003gk} using the Cayley-Hamilton theorem, which also affects the details of the stout force recursion.
The characteristic polynomial of a $3 \times 3$ matrix $A$ reads as follows:
\begin{equation}
    p_A(\lambda) = \lambda^3 - \tr[A] \lambda^2 + \frac{1}{2} \left( \tr[A]^2 - \tr[A^2] \right) \lambda - \det(A).
\end{equation}
For a traceless Hermitian matrix $Q$, this reduces to
\begin{equation}
    p_Q(\lambda) = \lambda^3 - \frac{1}{2} \tr[Q^2] \lambda - \det(Q).
\end{equation}
According to the Cayley-Hamilton theorem, $Q$ satisfies its own characteristic equation:
\begin{equation}
    Q^3 - c_1 Q - c_0 I = 0,
    \label{eq:Cayley-Hamilton_Q}
\end{equation}
where we again follow the notation in \cite{Morningstar:2003gk}:
\begin{align}
    c_0 &= \det(Q) = \frac{1}{3} \tr[Q^3], \\
    c_1 &= \frac{1}{2} \tr[Q^2] \geq 0.
\end{align}
The structure of $Q$ restricts the range of $c_0$:
\begin{equation}
    c_0^{\mathrm{max}} = -c_0^{\mathrm{min}} = 2 \Bigl( \frac{c_1}{3} \Bigr)^{3/2}.
    \label{eq:cayley-Hamilton_c0max}
\end{equation}
For analytic functions, \Cref{eq:Cayley-Hamilton_Q} may therefore be used to reduce the series representation of the function to a polynomial of degree $2$ in the matrix itself:
\begin{equation}
    f(Q) = f_2 Q^2 + f_1 Q + f_0.
\end{equation}
More generally, analytic functions of $n \times n$ matrices can be reduced to polynomials of degree $n - 1$ using the same strategy.
The complex coefficients $f_j$ may be expressed in terms of complex auxiliary functions $h_j$ and real variables $u$ and $w$ that are closely related to the eigenvalues of $Q$:
\begin{equation}
    f_j = \frac{h_j}{9u^2 - w^2},
    \label{eq:Cayley-Hamilton_f_j}
\end{equation}
where
\begin{align}
    u      &= \sqrt{\frac{c_1}{3}} \cos(\frac{\theta}{3}), \\
    w      &= \sqrt{c_1} \sin(\frac{\theta}{3}),           \\
    \theta &= \arccos(\frac{c_0}{c_0^{\mathrm{max}}}).
    \label{eq:Cayley-Hamilton_theta}
\end{align}
The $h_j$ in turn are defined as
\begin{align}
    \begin{split}
        h_0 =& \; (u^2 - w^2)e^{2i u}+e^{-iu} \bigl(8u^2 \cos(w) \\
             &+ 2iu (3u^2 + w^2) \xi_0(w) \bigr),
    \end{split}                                                  \\
    \begin{split}
        h_1 =& \; 2u e^{2i u} - e^{-i u} \bigl(2u \cos(w)        \\
             &-i (3u^2 - w^2) \xi_0(w) \bigr),
    \end{split}                                                  \\
    \begin{split}
        h_2 =& \; e^{2iu} - e^{-iu} \bigl(\cos(w) + 3iu \xi_0(w) \bigr),
    \end{split}
\end{align}
with
\begin{align}
    \xi_0(w) &= \sinc(w) = \frac{\sin(w)}{w},
    \label{eq:xi_0} \\
    \xi_1(w) &= \frac{\sinc'(w)}{w} = \frac{\cos(w)}{w^2} - \frac{\sin(w)}{w^3}.
    \label{eq:xi_1}
\end{align}
For small values of $w$, the direct evaluation of the expressions in \Cref{eq:xi_0} and \Cref{eq:xi_1} may suffer from numerical inaccuracies. To ensure sufficient precision, we use sixth-order Taylor expansions below certain thresholds:
\begin{align}
    \xi_0(w) &=
    \begin{cases}
        \mathrlap{1 - \frac{w^2}{6} \bigl(1 - \frac{w^2}{20} (1 - \frac{w^2}{42})\bigr)}
        \hphantom{-\frac{1}{3} + \frac{w^2}{30} \bigl(1 - \frac{w^2}{28} (1 - \frac{w^2}{54})\bigr)} & \abs{w} \leq \xi_{0, \mathrm{thr}}, \\
        \frac{\sin(w)}{w} & \abs{w} > \xi_{0, \mathrm{thr}}.
    \end{cases}                                                                                                                            \\
    \xi_1(w) &=
    \begin{cases}
        -\frac{1}{3} + \frac{w^2}{30} \bigl(1 - \frac{w^2}{28} (1 - \frac{w^2}{54})\bigr) & \abs{w} \leq \xi_{1, \mathrm{thr}},            \\
        \frac{\cos(w)}{w^2} - \frac{\sin(w)}{w^3} & \abs{w} > \xi_{1, \mathrm{thr}}.
    \end{cases}
\end{align}
The thresholds used in our simulations were determined empirically:
\begin{align}
    \xi_{0, \mathrm{thr}} &=
    \begin{cases}
        0.56\hphantom{0} & \text{in single precision,} \\
        0.05             & \text{in double precision.}
    \end{cases}                                        \\
    \xi_{1, \mathrm{thr}} &=
    \begin{cases}
        0.75  & \text{in single precision,}            \\
        0.115 & \text{in double precision.}
    \end{cases}
\end{align}
The stout force recursion also involves terms related to the derivative of the exponential function that appear in \Cref{eq:stout_force_gamma}. Specifically, the matrix polynomials $B_i$ are defined as
\begin{equation}
    B_i = b_{i2} Q^2 + b_{i1} Q + b_{i0}.
\end{equation}
The coefficients $b_{ij}$ are given by
\begin{align}
    b_{1j} &= \frac{2 u r_{j}^{(1)} + (3u^2 - w^2) r_{j}^{(2)} - 2 (15u^2 + w^2) f_j}{2(9u^2 - w^2)^2}, \label{eq:Cayley-Hamilton_b_1j} \\ 
    b_{2j} &= \frac{r_{j}^{(1)} - 3u r_{j}^{(2)} - 24u f_j}{2(9u^2 - w^2)^2},                           \label{eq:Cayley-Hamilton_b_2j}
\end{align}
and the auxiliary quantities $r_j^{(i)}$ are defined as
\begin{align}
    \begin{split}
        r^{(1)}_0 =& \; 2\Bigl(u + i(u^2 - w^2) \Bigr)e^{2iu}                             \\
                   &+ 2e^{-iu} \Bigl(4u \bigl(2 - iu \bigr)\cos(w)                        \\
                   &+ i \big(9u^2 + w^2- iu (3u^2 + w^2) \big) \xi_0(w) \Bigr),
    \end{split}                                                                           \\
    \begin{split}
        r^{(1)}_1 =& \; 2(1 + 2iu) e^{2iu} + e^{-iu} \Bigl(-2(1 - iu) \cos(w)             \\
                   &+ i \big(6u + i(w^2 - 3u^2) \big)\xi_0(w) \Bigr),
    \end{split}                                                                           \\
    \begin{split}
        r^{(1)}_2 =& \; 2i e^{2iu} + i e^{-iu} \Bigl( \cos(w) - 3(1 - iu)  \xi_0(w) \Bigr),
    \end{split}                                                                           \\
    \begin{split}
        r^{(2)}_0 =& \; -2e^{2iu} + 2i u e^{-iu} \Bigl( \cos(w)                           \\
                   &+ (1 + 4iu) \xi_0(w) + 3u^2 \xi_1(w) \Bigr),
    \end{split}                                                                           \\
    r^{(2)}_1 =& \; -i e^{-iu} \Bigl( \cos(w) + (1 + 2iu) \xi_0(w) -3u^2 \xi_1(w) \Bigr), \\
    r^{(2)}_2 =& \; e^{-iu} \Bigl( \xi_0(w) -3i u \xi_1(w) \Bigr).
\end{align}
In addition to the previously discussed functions $\xi_0(w)$ and $\xi_1(w)$, numerical instabilities may also be caused by a small denominator appearing in \Cref{eq:Cayley-Hamilton_f_j} and \Cref{eq:Cayley-Hamilton_b_1j,eq:Cayley-Hamilton_b_2j}. In particular, the expressions become problematic when $w \rightarrow 3u$, as either $c_{0} \rightarrow c_0^{\mathrm{min}}$ or $c_1 \rightarrow 0$. The former case can be circumvented by using the following symmetry relations of $f_j$ and $b_{ij}$ with respect to the sign of $c_0$:
\begin{align}
    f_{j}(-c_{0}, c_{1}) &= (-1)^{j} f_{j}^{*}(c_{0}, c_{1}), \\
    b_{ij}(-c_0, c_1)    &= (-1)^{i + j + 1} b_{ij}^{*}(c_0, c_1).
\end{align}
By only working with positive $c_0$, the range of $\theta$ is restricted to $\theta \in [0, \pi/2]$, which in turn implies that the expression $9u^2-w^2$ will lie in the interval $[2, 3]$.

The latter case $c_1 \rightarrow 0$ implies that $c_0^{\mathrm{max}} \rightarrow 0$, which can cause problems during the division in the argument of the $\arccos$ in \Cref{eq:Cayley-Hamilton_theta}. In practice, this can be handled by explicitly setting
\begin{gather}
    u   = w = 0, \\
    f_j = h_j =
    \begin{cases}
        1 & j = 0, \\
        0 & \text{else}.
    \end{cases}
\end{gather}
for very small values of $c_0^{\mathrm{max}}$ (namely, values smaller than the smallest representable positive normalized floating point number). In particular, this also covers the case of the trivial gauge configuration where $Q = 0$ and all three eigenvalues of the matrix will be degenerate and exactly equal to $0$.

Finally, problems appear if the argument of the $\arccos$ in \Cref{eq:Cayley-Hamilton_theta} exceeds 1 due to numerical inaccuracies. Since the argument is guaranteed to be bounded by 1 in exact arithmetic, this problem can be solved by rounding the argument down to 1.

\section{Exact solution of 2-dimensional U(1) gauge theory}
\label{appendix:U1_exact_solution}
In 2-dimensional U(1) gauge theory, exact results for expectation values of many observables are known analytically \cite{Kovacs:1995nn, Elser:2001pe, Bonati:2019ylr, Albandea:2021lvl}. A detailed derivation for Wilson loops in the aforementioned theory with both open and periodic boundary conditions can be found in Appendix A.3 of \cite{Kanwar:2021wzm}, of which we summarize the main results. For convenience, we introduce the notation
\begin{equation}
    R_{n, m}(\beta) = \frac{I_n(\beta)}{I_m(\beta)},
\end{equation}
where $I_n(\beta)$ is the $n$th modified Bessel function of the first kind.

For open boundary conditions in 2 dimensions, all plaquettes decouple and can be integrated independently. In particular, a consequence of this is the complete absence of finite volume effects, independent of the gauge group. Expectation values of Wilson loops with area $A$ (the exact shape does not matter due to the Abelian nature of the gauge group) are given by
\begin{equation}
    \langle \mathcal{W}_{A} \rangle_{\mathrm{open}} = \biggl( \frac{I_1(\beta)}{I_0(\beta)} \biggr)^A = \bigl( R_{1, 0}(\beta) \bigr)^A.
\end{equation}
Evidently, the plaquette expectation value is simply given by
\begin{equation}
    \langle P \rangle_{\mathrm{open}} = \frac{I_1(\beta)}{I_0(\beta)} = R_{1, 0}(\beta).
\end{equation}
When periodic boundary conditions are imposed, the theory is not completely trivial, as additional constraints are imposed by the periodicity. In that case, the expectation values of Wilson loops of area $A$ on a square lattice of size $L \times L$ can be expressed in terms of sums over irreducible characters of the gauge group:
\begin{equation}
    \langle \mathcal{W}_{A} \rangle_{\mathrm{periodic}} = \frac{\sum_r \bigl( I_r(\beta) \bigr)^{L^2 - A} \bigl( I_{r + 1}(\beta) \bigr)^A}{\sum\limits_r \bigl( I_r(\beta) \bigr)^{L^2}},
\end{equation}
where the index $r \in \mathbb{Z}$ runs over the irreducible representations of U(1). The direct evaluation of this expression is problematic for fine lattice spacings and larger volumes, since the power of the Bessel functions quickly exceeds the representable range for floating point numbers. This problem can be circumvented by slightly rewriting the expression and only working with ratios of Bessel functions:
\begin{equation}
    \langle \mathcal{W}_{A} \rangle_{\mathrm{periodic}} = \frac{\sum_r \bigl( R_{r, 0}(\beta) \bigr)^{L^2} \bigl( R_{r + 1, r}(\beta) \bigr)^A}{\sum\limits_r \bigl( R_{r, 0}(\beta) \bigr)^{L^2}}.
\end{equation}
For the plaquette, this simplifies to
\begin{equation}
    \langle P \rangle_{\mathrm{periodic}} = \frac{\sum_r \bigl( R_{r, 0}(\beta) \bigr)^{L^2} R_{r + 1, r}(\beta)}{\sum\limits_r \bigl( R_{r, 0}(\beta) \bigr)^{L^2}}.
\end{equation}
Finally, the (unnormalized) topological susceptibility is given by \cite{Albandea:2021lvl}
\begin{equation}
\begin{aligned}
    \langle Q^2 \rangle_{\mathrm{periodic}} = -V &\frac{\sum_r A_r(\beta) \bigl( I_r(\beta) \bigr)^{V-1}}{\sum\limits_r \bigl( I_r(\beta) \bigr)^{V}} \\
    - (V^2 - V) &\frac{\sum_r \bigl( B_r(\beta) \bigr)^2 \bigl( I_r(\beta) \bigr)^{V-2}}{\sum\limits_r \bigl( I_r(\beta) \bigr)^{V}},
\end{aligned}
\end{equation}
where
\begin{align}
    A_r(\beta) &= -\frac{1}{2 \pi} \int\limits_{-\pi}^{\pi} \mathrm{d}\phi \, \Bigl( \frac{\phi}{2 \pi} \Bigr)^2 e^{ir \phi + \beta \cos\phi}, \\
    B_r(\beta) &=  \frac{i}{2 \pi} \int\limits_{-\pi}^{\pi} \mathrm{d}\phi \,        \frac{\phi}{2 \pi}          e^{ir \phi + \beta \cos\phi}.
\end{align}
Again, it is advantageous from a numerical standpoint to work with ratios of Bessel functions:
\begin{equation}
\begin{aligned}
    \langle Q^2 \rangle_{\mathrm{periodic}} = -V &\frac{\sum_r A_r(\beta) \bigl( R_{r, 0}(\beta) \bigr)^{V-1}}{I_{0}(\beta) \sum\limits_r \bigl( R_{r, 0}(\beta) \bigr)^{V}} \\
    - (V^2 - V) &\frac{\sum_r \bigl( B_r(\beta) \bigr)^2 \bigl( R_{r, 0}(\beta) \bigr)^{V-2}}{\bigl( I_{0}(\beta) \bigr)^2 \sum\limits_r \bigl( R_{r, 0}(\beta) \bigr)^{V}}.
\end{aligned}
\end{equation}
All of the sums in the previous expressions converge rapidly; in fact, for our parameters ($V=32^2$, $\beta = 12.8$), only considering the leading terms already gives the correct results up to machine precision, which are indistinguishable from the results obtained for open boundary conditions for the Wilson loops.
\bibliography{literature}

\end{document}